\documentclass[twocolumn]{aastex701}
\shorttitle{Recovery of properties from infrared observations}
\shortauthors{Farrah et al.}

\begin{document}

\title{How accurately can obscured galaxy luminosities be measured using spectral energy distribution fitting of near- through far-infrared observations?}

\correspondingauthor{Duncan Farrah}
\email{dfarrah@hawaii.edu}

\author[0000-0003-1748-2010]{Duncan~Farrah}
\affiliation{Department of Physics and Astronomy, University of Hawai`i at M\=anoa, 2505 Correa Rd., Honolulu, HI, 96822, USA}
\affiliation{Institute for Astronomy, University of Hawai`i,  2680 Woodlawn Dr., Honolulu, HI, 96822, USA}
\email{dfarrah@hawaii.edu}

\author[0009-0002-6970-5247]{Kiana Ejercito}
\affiliation{Department of Physics, University of California, Santa Barbara, Santa Barbara, CA 93106, USA}
\email{Kiana@ucsb.edu}

\author[0000-0002-2612-4840]{Andreas Efstathiou}
\affiliation{School of Sciences, European University Cyprus, Diogenes Street, Engomi, 1516 Nicosia, Cyprus}
\email{A.Efstathiou@euc.ac.cy}

\author[0000-0003-1000-2547]{David Leisawitz}
\affiliation{Exoplanets and Stellar Astrophysics Laboratory, NASA Goddard Space Flight Center, Code 667, 8800 Greenbelt Rd., Greenbelt, MD, 20771, USA}
\email{david.t.leisawitz@nasa.gov}

\author[0000-0001-6970-7782]{Athena Engholm}
\affiliation{Department of Astronomy, University of Illinois Urbana-Champaign, Urbana, IL, USA}
\email{athena@illinios.edu}

\author[0000-0003-4702-7561]{Irene Shivaei}
\affiliation{Centro de Astrobiolog\'{i}a (CAB), CSIC-INTA, Carretera de Ajalvir km 4, Torrej\'{o}n de Ardoz, 28850, Madrid, Spain}
\email{ishivaei@cab.inta-csic.es}

 \author[0000-0001-9139-2342]{Matteo Bonato}
\affiliation{INAF--Istituto di Radioastronomia and Italian ALMA Regional Centre, Via Gobetti 101, Bologna, Italy, I-40129}
\email{matteo.bonato@inaf.it}

 \author[0000-0002-9548-5033]{David L. Clements}
\affiliation{Imperial College London}
\email{d.clements@imperial.ac.uk}

\author[0000-0003-0624-3276]{Sara Petty}
\affiliation{NorthWest Research Associates, 3380 Mitchell Ln., Boulder, CO 80301, USA}
\email{spetty@nwra.com}

\author[0000-0002-5206-5880]{Lura. K. Pitchford}
\affiliation{Department of Physics and Astronomy, Texas A\&M University, College Station, TX, USA} 
\affiliation{George \& Cynthia Woods Mitchell Institute for Fundamental Physics and Astronomy, Texas A\&M University, College Station, TX, USA}
\email{kpitchford@tamu.edu}

\author[0009-0004-6200-0919]{Charalambia Varnava}
\affiliation{School of Sciences, European University Cyprus, Diogenes Street, Engomi, 1516 Nicosia, Cyprus}
\email{varnava.haris@gmail.com}
 
\author[0000-0002-9149-2973]{Jose Afonso}
\affiliation{Instituto de Astrofísica e Ciências do Espaço, Universidade de Lisboa, OAL, Tapada da Ajuda, PT1349-018 Lisbon, Portugal}
\affiliation{Departamento de Física, Faculdade de Ciências, Universidade de Lisboa, Edifício C8, Campo Grande, PT1749-016 Lisbon}
\email{jafonso@ciencias.ulisboa.pt}

\author[0000-0002-5836-4056]{Carlotta Gruppioni}
\affiliation{Istituto Nazionale di Astrofisica - Osservatorio di Astrofisica e Scienza dello Spazio, via Gobetti 93/3, 40129 Bologna, Italy}
\email{carlotta.gruppioni@inaf.it}

\author[0000-0003-0917-9636]{Evanthia Hatziminaoglou}
\affiliation{European Southern Observatory, Karl-Schwarzschild-Str. 2, 85748 Garching bei München, Germany}
\affiliation{Instituto de Astrofisica de Canarias (IAC), E-38205 La Laguna, Tenerife, Spain}
\affiliation{Universidad de La Laguna, Opto. Astrofisica, E-38206 La Laguna, Tenerife, Spain}
\email{ehatzimi@eso.org}

\author[0000-0002-8732-6980]{Andrew Hoffman}
\affiliation{Institute for Astronomy, University of Hawai`i,  2680 Woodlawn Dr., Honolulu, HI, 96822, USA}
\email{amho@hawaii.edu} 

\author[0000-0002-3032-1783]{Mark Lacy}
\affiliation{National Radio Astronomy Observatory, Charlottesville, VA, USA}
\email{mlacy@nrao.edu}

\author[0000-0003-3017-9577]{Brenda C. Matthews}
\affiliation{Herzberg Astronomy \& Astrophysics Research Centre, National Research Council of Canada, 5071 West Saanich Road, Victoria, BC, Canada}
\affiliation{Department of Physics \& Astronomy, University of Victoria, 3800 Finnerty Rd, Victoria, BC, Canada}
\email{bcmatthews.herzberg@gmail.com}

  \author[0000-0001-9540-9121]{Conor Nixon}
\affiliation{Planetary Systems Laboratory, NASA Goddard Space Flight Center, Greenbelt, MD, USA}
\email{conor.a.nixon@nasa.gov}

 \author[0000-0001-6139-649X]{Chris Pearson}
\affiliation{RAL Space, STFC Rutherford Appleton Laboratory, Didcot, Oxfordshire OX11 0QX, UK}
\affiliation{The Open University, Milton Keynes MK7 6AA, UK}
\affiliation{Oxford Astrophysics, University of Oxford, Keble Rd, Oxford OX1 3RH, UK}
\email{chris.pearson@stfc.ac.uk}

\author[0000-0001-9701-5660]{Berke Vow Ricketti}
\affiliation{Disruptive Space Technology Centre, RAL Space, STFC-Rutherford Appleton Laboratory, Didcot, UK}
\email{berke.ricketti@stfc.ac.uk }

\author[0000-0001-6854-7545]{Dimitra Rigopoulou}
\affiliation{Department of Physics, University of Oxford, Keble Road, Oxford OX1 3RH, UK}
\email{Dimitra.Rigopoulou@physics.ox.ac.uk}
 
\author[0000-0003-0796-362X]{Loren Robinson}
\affiliation{Department of Astronomy, University of Wisconsin-Madison, 475 N. Charter St., Madison, WI 53706, USA}
\email{ljrobinson4@wisc.edu}

\author[0000-0002-9941-2077]{Locke D. Spencer}
\affiliation{Institute for Space Imaging Science, Dept. of Physics \& Astronomy, University of Lethbridge, Alberta, Canada}
\email{Locke.Spencer@uLeth.ca}

\author[0000-0002-6736-9158]{Lingyu Wang}
\affiliation{SRON Netherlands Institute for Space Research}
\affiliation{University of Groningen}
\email{l.wang@sron.nll}

\author[0000-0002-1233-9998]{David B. Sanders}
\affiliation{Institute for Astronomy, University of Hawai`i,  2680 Woodlawn Dr., Honolulu, HI, 96822, USA}
\email{sanders@ifa.hawaii.edu}

\author[0000-0002-8552-158X]{Gerard van Belle}
\affiliation{Lowell Observatory, 1400 West Mars Hill Road, Flagstaff, AZ 86001, USA}
\email{gerard@lowell.edu}

\begin{abstract}
Infrared-luminous galaxies are important sites of stellar and black hole mass assembly at most redshifts. Their luminosities are often estimated by fitting spectral energy distribution (SED) models to near- to far-infrared data, but the dependence of these estimates on the data used is not well-understood.  Here, using observations simulated from a well-studied local sample, we compare the effects of wavelength coverage, signal-to-noise (S/N), flux calibration, angular resolution, and  redshift on the recovery of starburst, AGN, and host luminosities. We show that the most important factors are wavelength coverage that spans the peak in a SED, with dense wavelength sampling. Such observations recover starburst and AGN infrared luminosities with systematic bias below $20\%$.  Starburst luminosities are best recovered with far-infrared observations while AGN luminosities are best recovered with near- and mid-infrared observations, though the recovery of both are enhanced with near/mid-infrared, and far-infrared observations, respectively. Host luminosities are best recovered with near/far-infrared observations, but are usually biased low, by $\gtrsim20\%$. The recovery of starburst and AGN luminosity is enhanced by observing at high angular resolution. Starburst-dominated systems show more biased recovery of luminosities than do AGN-dominated systems.  As redshift increases, far-infrared observations become more capable, and mid-infrared observations less capable, at recovering luminosities. Our results highlight the transformative power of a far-infrared instrument with dense wavelength coverage from tens to hundreds of microns for studying infrared-luminous galaxies. We tabulate estimates of systematic bias and random error for use with JWST and other observatories. 
\end{abstract}


\section{Introduction}
Infrared-luminous galaxies are systems harboring brief ($\sim$ tens of Myr) intense phases of star formation and/or supermassive black hole (SMBH) accretion that are at least moderately obscured, leading to high infrared to optical luminosity ratios.  These systems are rare in the nearby Universe, but rise rapidly in number density with redshift, such that, by $z\sim1$, they harbor a significant fraction of the comoving star formation rate and SMBH accretion rate densities. They have also been linked to processes such as the merger-driven transformation of disks to spheroids, and AGN and/or star formation based feedback.  Reviews of their properties can be found in  \citet{sm96,blain02,lon06,cas14,perez21}.

Estimating the starburst, AGN, and host luminosities of infrared-luminous galaxies has been an observational goal since their discovery, and remains so in the era of JWST. The most direct method for estimating these luminosities, and the one that many other methods are calibrated on, is to  measure the shape of the ultraviolet through millimeter SEDs, and then use either energy balance arguments or radiative transfer modelling to decompose the SED into its starburst, AGN, and host components \citep{dacunha08,chevallard16,boquien19,smart24}. SED fitting is usually performed using photometric or low to moderate resolution spectroscopic ($R\lesssim200$) data, and constrains the fits based on the shape of the continuum, and broad spectral features such as the polycyclic aromatic hydrocarbon (PAH) features. This approach has been used in studies at all redshifts  \citep[e.g.][]{farrah03,farrah16,hatz08,hatz09,lon15,herr17,mat18,kool20,efs22,yamada23,spinoglio24,varna24,varna25a}.

\begin{deluxetable}{lcccccc}
\tablecaption{The input sample to our simulations. The columns give their IRAS names, redshifts, their assumed ``true" total and component rest-frame $1-1000\mu$m luminosities, and the PAH EW values used to divide the sample into starburst and AGN-dominated sub-samples (\S\ref{sec:sample_selection}.) 
\label{tbl:sample}}
\tablehead{
\colhead{Name} & 
\colhead{Redshift} & 
\colhead{$L^{True}_{Tot}$} & 
\colhead{$L^{True}_{Sb}$} & 
\colhead{$L^{True}_{AGN}$} & 
\colhead{$L^{True}_{H}$} & 
\colhead{$W_{6.2}$}  \\
\colhead{} & 
\colhead{} & 
\multicolumn{4}{c}{$\log_{10}(L_{\odot})$}& 
\colhead{$\mu$m}  
}
\startdata
IRAS 00188-0856  &  $ 0.128$  &  $ 12.35$  &  $ 12.27$  &  $ 11.45$  &  $ 11.03$  &  $ 0.102$ \\ 
IRAS 00397-1312  &  $ 0.262$  &  $ 12.97$  &  $ 12.85$  &  $ 12.32$  &  $ 11.19$  &  $ 0.036$ \\ 
IRAS 01003-2238  &  $ 0.118$  &  $ 12.20$  &  $ 11.88$  &  $ 11.91$  &  $ 10.36$  &  $ 0.055$ \\ 
IRAS 01572+0009 (Mrk 1014) &  $ 0.163$  &  $ 12.53$  &  $ 12.39$  &  $ 11.90$  &  $ 11.27$  &  $ 0.064$ \\ 
IRAS 03158+4227  &  $ 0.134$  &  $ 12.61$  &  $ 12.42$  &  $ 12.14$  &  $ 10.89$  &  $ 0.101$ \\ 
IRAS 03521+0028  &  $ 0.152$  &  $ 12.38$  &  $ 12.33$  &  $ 11.41$  &  $ 10.05$  &  $ 0.348$ \\ 
IRAS 05189-2524  &  $ 0.043$  &  $ 12.14$  &  $ 11.89$  &  $ 11.74$  &  $ 10.84$  &  $ 0.043$ \\ 
IRAS 06035-7102  &  $ 0.079$  &  $ 12.21$  &  $ 12.08$  &  $ 11.32$  &  $ 11.26$  &  $ 0.105$ \\ 
IRAS 06206-6315  &  $ 0.092$  &  $ 12.17$  &  $ 12.07$  &  $ 11.23$  &  $ 11.08$  &  $ 0.213$ \\ 
IRAS 07598+6508  &  $ 0.148$  &  $ 12.62$  &  $ 12.34$  &  $ 12.30$  &  $ 9.56$  &  $ 0.011$ \\ 
IRAS 08311-2459  &  $ 0.100$  &  $ 12.47$  &  $ 12.35$  &  $ 11.75$  &  $ 11.02$  &  $ 0.139$ \\ 
IRAS 08572+3915  &  $ 0.058$  &  $ 12.17$  &  $ 11.82$  &  $ 11.89$  &  $ 10.27$  &  $ <0.003$ \\ 
IRAS 09022-3615  &  $ 0.060$  &  $ 12.23$  &  $ 12.15$  &  $ 11.41$  &  $ 10.56$  &  $ 0.148$ \\ 
IRAS 09320+6134 (UGC 5101)  &  $ 0.039$  &  $ 12.01$  &  $ 11.93$  &  $ 10.94$  &  $ 10.97$  &  $ 0.250$ \\
IRAS 10378+1109  &  $ 0.136$  &  $ 12.24$  &  $ 11.91$  &  $ 11.92$  &  $ 10.95$  &  $ 0.105$ \\ 
IRAS 10565+2448  &  $ 0.043$  &  $ 12.05$  &  $ 11.97$  &  $ 10.97$  &  $ 10.97$  &  $ 0.635$ \\ 
IRAS 11095-0238  &  $ 0.107$  &  $ 12.19$  &  $ 11.77$  &  $ 11.98$  &  $ 10.27$  &  $ 0.054$ \\ 
IRAS 12071-0444  &  $ 0.128$  &  $ 12.34$  &  $ 12.08$  &  $ 11.93$  &  $ 11.21$  &  $ 0.118$ \\ 
IRAS 12540+5708 (Mrk 231)  &  $ 0.042$  &  $ 12.54$  &  $ 12.28$  &  $ 12.17$  &  $ 11.02$  &  $ 0.021$ \\ 
IRAS 13120-5453  &  $ 0.031$  &  $ 12.24$  &  $ 12.18$  &  $ 10.99$  &  $ 11.14$  &  $ 0.528$ \\ 
IRAS 13428+5608 (Mrk 273) &  $ 0.037$  &  $ 12.11$  &  $ 12.04$  &  $ 11.24$  &  $ 10.56$  &  $ 0.190$ \\ 
IRAS 13451+1232 (4C 12.50) &  $ 0.122$  &  $ 12.28$  &  $ 11.54$  &  $ 12.12$  &  $ 11.38$  &  $ 0.011$ \\ 
IRAS 13536+1836 (Mrk 463)  &  $ 0.049$  &  $ 11.88$  &  $ 11.36$  &  $ 11.70$  &  $ 10.59$  &  $<0.002$ \\ 
IRAS 14348-1447  &  $ 0.083$  &  $ 12.30$  &  $ 12.12$  &  $ 11.59$  &  $ 11.43$  &  $ 0.249$ \\ 
IRAS 14378-3651  &  $ 0.068$  &  $ 12.05$  &  $ 11.98$  &  $ 11.15$  &  $ 10.63$  &  $ 0.445$ \\ 
IRAS 15250+3609  &  $ 0.055$  &  $ 12.02$  &  $ 11.82$  &  $ 11.57$  &  $ 10.40$  &  $ 0.044$ \\ 
IRAS 15327+2340 (Arp 220) &  $ 0.018$  &  $ 12.03$  &  $ 11.96$  &  $ 11.06$  &  $ 10.79$  &  $ 0.272$ \\ 
IRAS 15462-0450  &  $ 0.100$  &  $ 12.10$  &  $ 11.98$  &  $ 11.40$  &  $ 10.61$  &  $ 0.079$ \\ 
IRAS 16090-0139  &  $ 0.134$  &  $ 12.48$  &  $ 12.43$  &  $ 11.42$  &  $ 10.81$  &  $ 0.093$ \\ 
IRAS 16504+0228 (NGC 6240) &  $ 0.024$  &  $ 11.83$  &  $ 11.62$  &  $ 11.09$  &  $ 11.14$  &  $ 0.378$ \\ 
IRAS 17208-0014  &  $ 0.043$  &  $ 12.35$  &  $ 12.32$  &  $ 10.78$  &  $ 10.83$  &  $ 0.457$ \\ 
IRAS 19254-7245 (SuperAntena)  &  $ 0.062$  &  $ 12.12$  &  $ 11.85$  &  $ 11.44$  &  $ 11.51$  &  $ 0.074$ \\ 
IRAS 19297-0406  &  $ 0.086$  &  $ 12.40$  &  $ 12.29$  &  $ 11.53$  &  $ 11.29$  &  $ 0.436$ \\ 
IRAS 20087-0308  &  $ 0.106$  &  $ 12.43$  &  $ 12.40$  &  $ 10.98$  &  $ 10.99$  &  $ 0.352$ \\ 
IRAS 20100-4156  &  $ 0.130$  &  $ 12.51$  &  $ 12.42$  &  $ 11.71$  &  $ 10.78$  &  $ 0.139$ \\ 
IRAS 20414-1651  &  $ 0.087$  &  $ 12.16$  &  $ 12.10$  &  $ 11.16$  &  $ 10.81$  &  $ 0.571$ \\ 
IRAS 20551-4250  &  $ 0.043$  &  $ 12.01$  &  $ 11.79$  &  $ 11.49$  &  $ 10.93$  &  $ 0.111$ \\ 
IRAS 22491-1808  &  $ 0.078$  &  $ 12.07$  &  $ 11.87$  &  $ 11.55$  &  $ 10.91$  &  $ 0.403$ \\ 
IRAS 23128-5919  &  $ 0.045$  &  $ 11.87$  &  $ 11.75$  &  $ 11.12$  &  $ 10.64$  &  $ 0.364$ \\ 
IRAS 23230-6926  &  $ 0.107$  &  $ 12.16$  &  $ 12.03$  &  $ 11.53$  &  $ 10.36$  &  $ 0.312$ \\ 
IRAS 23253-5415  &  $ 0.130$  &  $ 12.31$  &  $ 12.05$  &  $ 11.76$  &  $ 11.53$  &  $ 0.288$ \\ 
IRAS 23365+3604  &  $ 0.064$  &  $ 12.16$  &  $ 12.06$  &  $ 11.42$  &  $ 10.58$  &  $ 0.401$ \\ 
\enddata
\end{deluxetable}

Understanding how systematic bias and random error (accuracy and precision, respectively) affect the results from SED fitting is essential to their reliable interpretation.  Furthermore, planning for future infrared observatories is enhanced by knowing which factors affect the reliability of SED fitting results \citep{farrah19}. Both accuracy and precision may depend on which data the SED models are fit to, since SED shapes can be degenerate, especially in the rest-frame near- and mid-infrared for starbursts vs. AGN, and the rest-frame near- and far-infrared for starbursts vs. their host galaxies.  Yet, this issue has not been thoroughly investigated.  There exist studies that examine how well galaxy properties can be recovered from specific survey designs \citep[e.g.][]{bonatto24,bisig24}, and which compare the results from different SED fitting codes \citep{huntsed19,pacifici23}. However, the dependence of recovered luminosities on the data used is less well-studied \citep[though see][]{magris15,leja19how,thorne23,sommo25}. 

In this work, we examine the issues that affect the accuracy and precision of starburst, AGN, and host luminosities recovered via SED fitting of infrared-luminous galaxies. We use a well-studied sample of infrared-luminous galaxies to generate mock observations for fiducial instruments. We then apply a standard SED fitting approach to those mock observations and compare the derived luminosities to their (assumed) true values. We define our input sample in \S\ref{sec:sample_selection}.  Using their SEDs as the ``true" SEDs, we describe the generation of mock observations for fiducial near- to far-infrared instruments that are representative of past, current, and future facilities in \S\ref{sec:instru} and \ref{sec:obsgen}. We describe the fits to these mock observations and the recovery of component luminosities in \S\ref{sec:obsrec}. We compare the recovered luminosities to their "true" values, and present estimates of the accuracy and precision of luminosity recovery for each instrument, in \S\ref{sec:res}. We present our discussion and conclusions in  \S\ref{sec:discCon}. Throughout, we adopt the same cosmology as in \citet{efs22}; a spatially flat universe with \mbox{$H_0 = 70$\,km\,s$^{-1}$\,Mpc$^{-1}$} and \mbox{$\Omega_{\Lambda} = 0.7$}. 
 
\begin{figure}
\begin{center}
\includegraphics[width=0.98\linewidth]{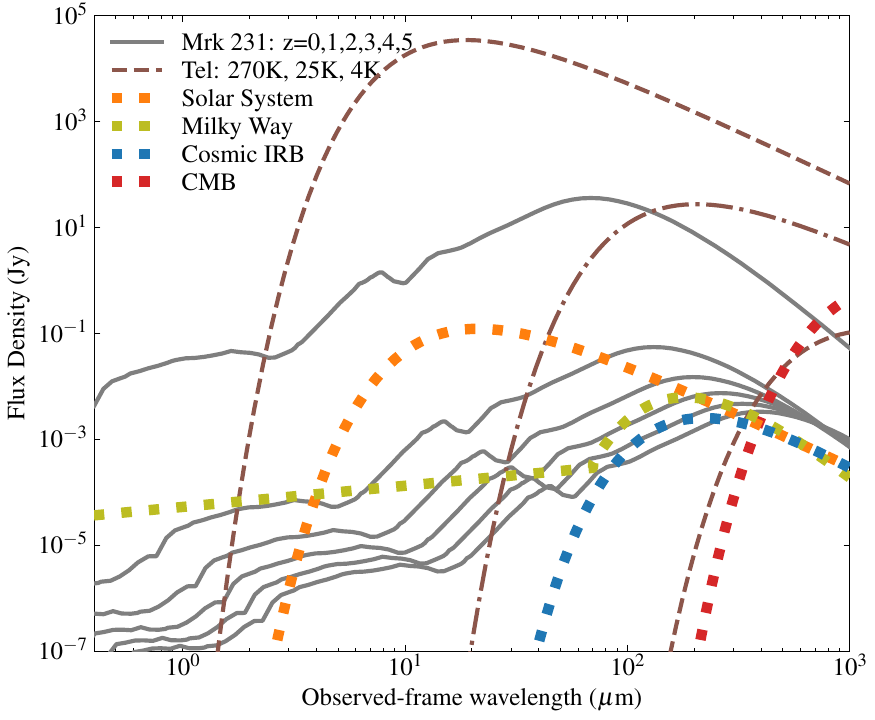} 
\caption{Comparison of telescope and astrophysical backgrounds in a $10\arcsec\times10\arcsec$ aperture, with Mrk 231 at $z=0,1,2,3,4,5$ (\S\ref{sec:obsgen}).  With $B(T)$ as the Planck function, the telescope backgrounds follow $\epsilon_{tel} B(T_{tel})$ with $\epsilon_{tel}=0.04$. Zodiacal light is modelled as $2\times10^{-7} B(T_{zod})$ with $T_{zod}=240\,$K \citep{leinert02}. Milky Way emission is modelled via $40 ( (\lambda)/(\lambda_{0})      )^{-\beta} 2\times10^{-7} B(T_{ISM})$  (with $T_{ism} = 17.53\,$K,  $\lambda_{0} = 23.05\,\mu$m, and $\beta = 1.55$) for wavelengths longward of $70\,\mu$m, and  $5\times10^{-23} ((\lambda)/(\lambda_{0}))^{0.4}$ at wavelengths shortward of $70\,\mu$m \citep{finkbein99}. For the cosmic infrared background, we adopt $1.6\times10^{-5}\left(\frac{\nu}{\nu_{0}}\right)^{(0.64)}  B(T_{IRB})$. The CMB is modelled as $B(T_{CMB})$, with $T_{CMB} = 2.718$\,K \citep{mather94}.}
\label{fig:backgrounds}
\end{center}
\end{figure}

\begin{deluxetable}{lcc}
\tablecaption{Simulated instruments, their wavelength ranges, and fiducial wavelengths for S/N calculations (\S\ref{sec:instru}). `Opt' corresponds to the Sloan $g$ and $r$ filters. For other instrument names: `N', `M', `F', and `S' correspond to near-infrared, mid-infrared, far-infrared, and sub-mm, respectively, and `s' and `p' correspond to spectroscopy and photometry. The NMp filters correspond to JWST F222W, F444W, F777W, F1500W, and F2200W. The Fp filters correspond to \textit{Herschel} PACS/SPIRE photometry. 
\label{tbl:instparams}}
\tablehead{
\colhead{Instrument} & 
\colhead{$\lambda$ ($\mu$m)} & 
\colhead{$\lambda_{fid}$ ($\mu$m)} \\
}
\startdata
Opt.\tablenotemark{a}    & 0.45, 0.70    & 0.45, 0.70    \\
\hline
Ns       & 1.4-4.8        & 2.0              \\
Ms       &  5.0-28.0  & 10.0    \\
NMp    & 2.2, 4.4, 7.7, 15, 22  & 2.2, 4.4, 7.7, 15, 22 \\
Fs        &  30-300   & 50    \\
Fp        &  70,160,250,350,500  & 70,160,250,350,500  \\
\hline
Sp\tablenotemark{b} &  450,850                  & 450,850  \\
\enddata
\tablenotetext{a}{Common to all simulations (\S\ref{sec:instru}).}
\tablenotetext{b}{Only used in multi-instrument simulations (\S\ref{sec:multi}).}
\end{deluxetable}

\begin{figure*}
\begin{center}
\includegraphics[width=0.98\linewidth]{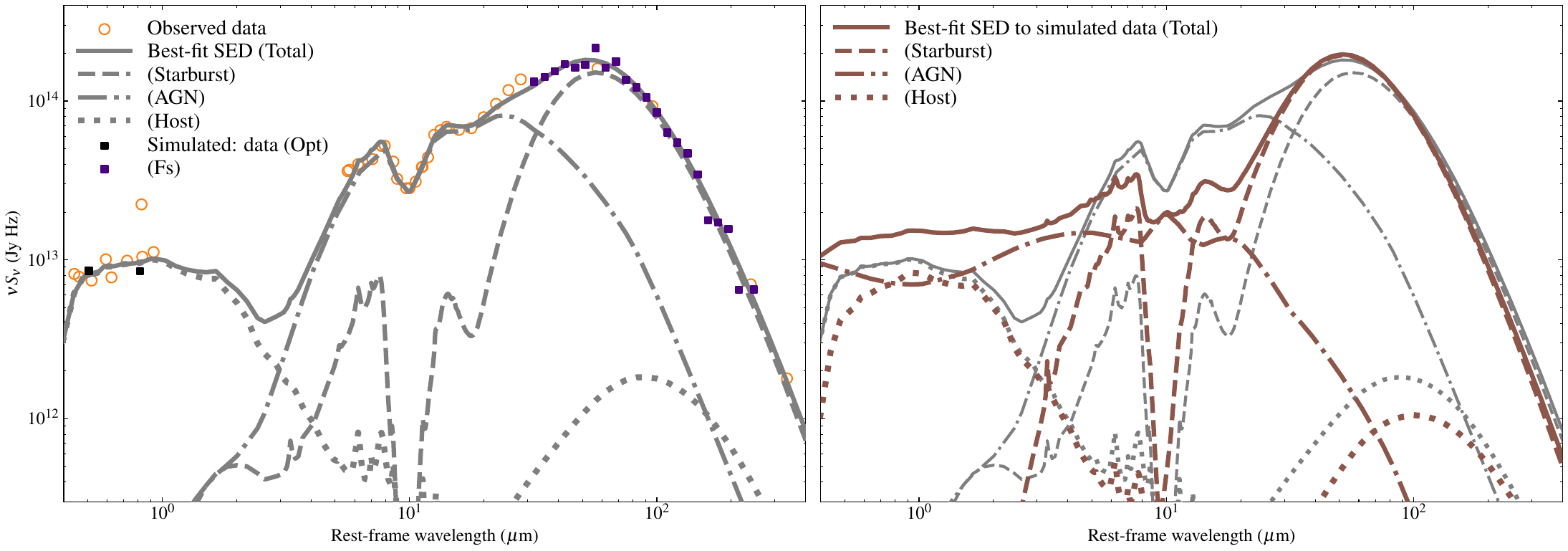} 
\caption{An example of an input SED, simulated data, and recovered fit (\S\ref{sec:obsrec}). The left panel shows the original observed data and best-fit SED for Mrk 231  \citep[figure A7 in][]{efs22}. Also shown are the simulated Opt and Fs data (Table \ref{tbl:instparams}) from this best-fit "truth" SED. The right panel repeats the "truth" SED from the left panel, along with the SED obtained from fitting to the simulated Opt and Fs data.  In this case the recovered starburst SED is well matched to the "truth" starburst SED, but the AGN and host components diverge significantly (\S\ref{sec:resnear}).}
\label{fig:examplesed}
\end{center}
\end{figure*}

\section{Input Sample}\label{sec:sample_selection}
As the input sample for our simulations, we adopt the 42 $z<0.27$ ultraluminous infrared galaxies (ULIRGs)  that were observed as part of the HERschel ULIRG Reference Survey \citep{far13,spoon13,pea16,cle18}.  This sample is an ideal input population for our study for three reasons.  First, the sample is an almost unbiased set of nearby infrared-luminous galaxies in which the starburst and/or AGN are much brighter than their host galaxies.  The selection is detailed in \citet{far13} and comprises nearly all known systems with $L_{IR}\gtrsim10^{12}$L$_{\odot}$ at $z<0.27$. The sample thus spans the full range of starburst to AGN dominance, and all four primary optical spectral classes (\ion{H}{2}, LINER, Sy2, Sy1). Second, their ultraviolet through millimeter SEDs have been comprehensively modelled using a radiative transfer approach  \citep{efs22,farrah22}, giving a library of SEDs, each with a starburst, AGN, and host component, from which to build our ``truth" observations.  The details are in \citet{efs22} but in brief: The starburst \citep{efstathiou09} is modelled as an exponentially decaying burst in which the free parameters include the initial optical depths of the star-forming clouds, the e-folding times of the starburst, and starburst age. The AGN model \citep{efsrr95,efstathiou95,efstathiou13} assumes a smooth, tapered disk in which the half-opening and inclination angles of the torus, the ratio of inner to outer radius, and equatorial optical depth, are the primary free parameters. The host model assumes a S\'{e}rsic profile with $n=4$, and a mixed distribution of stars and dust. The SEDs have a spectral resolution of $R\simeq200$ and include the PAH features between $3\mu$m and $13\mu$m, and the silicate dust features at $9.7\mu$m and $18\mu$m. They do not, however, include mid- or far-infrared fine structure lines. Third, a sample size of 42 is a reasonable compromise between having enough objects to assess the range in recovered luminosities, while allowing the fitting to complete using moderate computing resources. 

 We make one simplifying assumption to this sample. \citet{efs22} find that three of them show evidence for a ``polar" dust SED component - dust within the polar regions of the AGN at distances of a few parsecs. Polar dust is thus rare (though see \citealt{lyurieke22}), and less luminous than the starburst or AGN. So, we exclude the polar dust component from these three objects to give input SEDs of 42 objects whose components are their starbursts, AGN, and host galaxies. The sample and their `true' luminosities are presented in Table \ref{tbl:sample}.  Diagnostic plots for the sample can be found in \citet{efs22} and \citet{farrah22}. 
 
This choice of input sample places two qualifiers on our results. First, they are most relevant to objects whose starburst and/or AGN are at least a factor of five times as luminous as their hosts. Our results become less applicable to objects as this ratio declines. Second, our results are most relevant for objects which are drawn at random from a parent population that resembles our input sample (though we consider simulations using subsets of the input population that are starburst or AGN dominated in \S\ref{sec:paheffect}).

\section{Simulated Instruments}\label{sec:instru}
We consider observations from the following mock near- through far-infrared instruments:

\begin{itemize}

\item {\bfseries Ns}: a near-infrared spectrograph covering $1.5-5.0\mu$m. This  resembles spectrographs such as SpeX on the NASA Infrared Telescope Facility \citep{rayner03}, NIRS on Gemini \citep{elias06}, IRCS on Subaru \citep{tokunaga98}, the shorter wavelength modules within the IRS on \textit{Spitzer} \citep{houck04}, and NIRSpec on JWST \citep{jakobsen22}. 

\item {\bfseries Ms}: a mid-infrared spectrograph covering $5-28\mu$m. This   resembles instruments with coverage of both the N and Q bands, the longer wavelength modules within the IRS onboard \textit{Spitzer} and MIRI onboard JWST \citep{rieke23miri}. The combination of Ns and Ms resembles the SWS instrument onboard ISO \citep{degraauw98}. 

\item {\bfseries NMp}: photometric observations in five bands spanning near- to mid-infrared wavelengths. We adopt band centers and filter widths that resemble commonly-used filters on NIRCam \citep{riekem23} and MIRI on JWST, but the results should be applicable to any choice of filters that populate near- to mid-infrared wavelengths. 

\item {\bfseries Fs}: a far-infrared spectrograph covering $30-300\mu$m. This instrument is similar to those proposed for concept missions, including the FIRESS \citep{firess25} onboard PRIMA \citep{primaburg24,glenn25}, SAFARI on SPICA \citep{egami18}, SPIRIT \citep{spirit07,spice24}, and SALTUS \citep{saltus24,spilker24}. 

\item {\bfseries Fp}: Photometry in five far-infrared photometric bands, centered at $70, 160, 250, 350$, and $500\mu$m, chosen to match those within the PACS and SPIRE instruments onboard \textit{Herschel} \citep{griffin10,pogl10}. Fp also resembles future instruments such as Primager on PRIMA \citep{primager25}.

\end{itemize}

\noindent In addition, we consider the following optical and sub-millimeter instruments:

\begin{itemize}

\item {\bfseries Opt:} Since optical imaging in at least one band is ubiquitous, we assume that all simulated observations at any redshift include imaging in two optical bands.  We select the Sloan Digital Sky Survey (SDSS) $g$ and $r$ bands, which are almost identical to the $g$ and $r$ band filters on the Vera Rubin Observatory. 

\item {\bfseries Sp}: When considering observations with multiple instruments (\S\ref{sec:multi}) we include a sub-millimeter instrument that performs photometry at $450\mu$m and $850\mu$m.  This instrument resembles several ground-based facilities that observe through two low opacity windows in Earth's atmosphere, including ALMA, SCUBA2 \citep{holland13}, or AtLAST \citep{atlast25}. We do not consider observations with Sp on its own as, at most redshifts, it constrains only the slope of the Rayleigh-Jeans tail of the SED. 

\end{itemize}

\noindent  The parameters for each instrument are given in Table \ref{tbl:instparams}. We assume that all instruments have flat sensitivity curves.  This assumption is reasonable, since filters generally have fairly flat sensitivity profiles, and most modern spectrographs, including NIRSpec and MIRI, have sensitivities that vary by less than a factor of two across most of their wavelength ranges.  The spectral resolutions of many of these instruments are higher than the spectral resolution of the input SEDs, but since our aim is to examine the recovery of infrared luminosities from SED fitting, rather than measuring fine-structure lines, this should not affect our analysis. We note though that if structure in continua at $R \gg 200$ can constrain luminosities, then our analysis will not include this effect.

\begin{figure*}
\begin{center}
\includegraphics[width=0.98\linewidth]{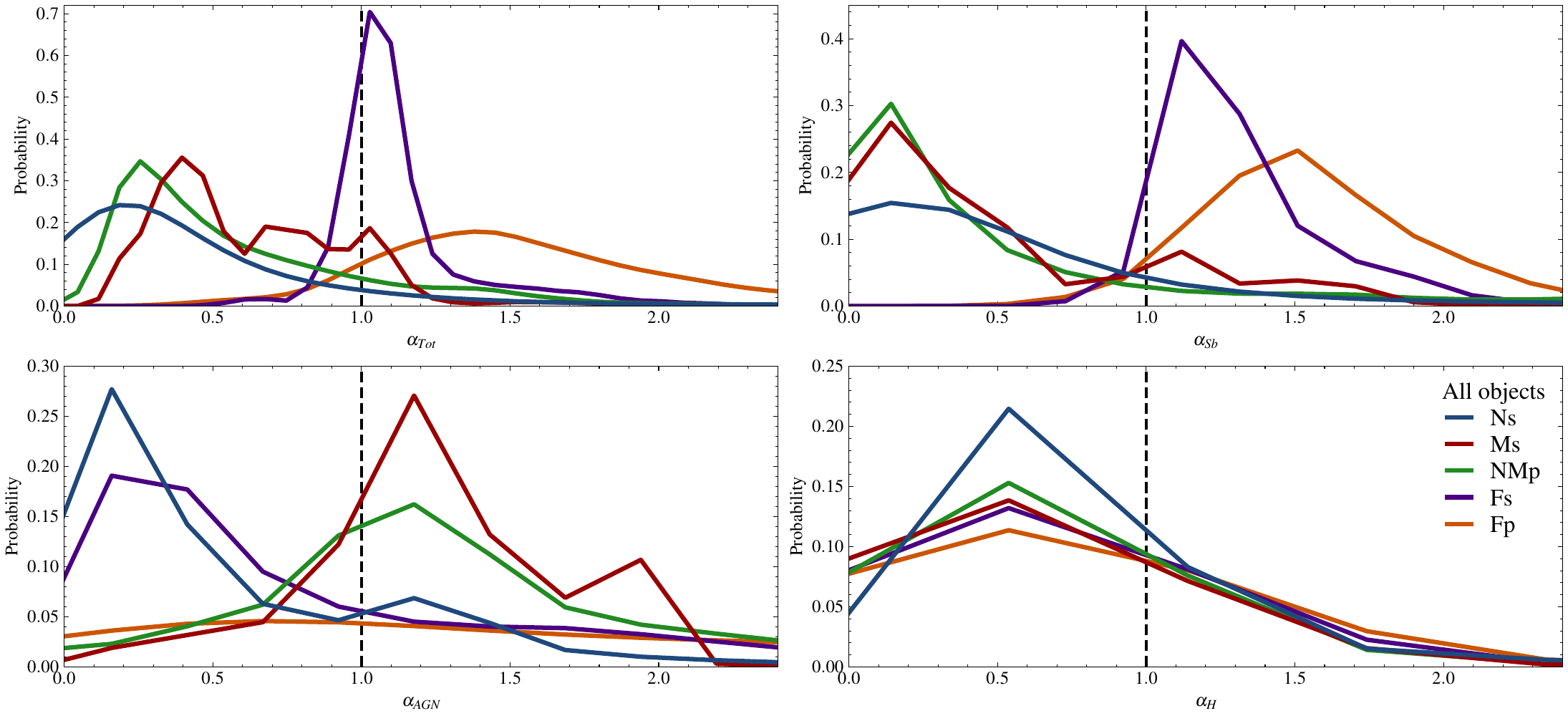} 
\caption{The fractional probability of recovering an observed-to-true luminosity ratio, $\alpha$ (Equation \ref{eq:alphadef}) at $z\sim0.1$, using observations with the Ns, Ms, NMp, Fs, and Fp instruments individually (\S\ref{sec:resnear}). The top left panel shows the recovery of total infrared luminosity ($L_{Tot}$), the top right panel shows the recovery of starburst luminosity ($L_{Sb}$), the bottom left panel shows the recovery of AGN luminosity ($L_{AGN}$), and the bottom right panel shows the recovery of host galaxy luminosity ($L_{H}$).}
\label{fig:localsingle}
\end{center}
\end{figure*}

\section{Observations Generation}\label{sec:obsgen}
Using the input sample SEDs (\S\ref{sec:sample_selection}) as the ``truth", we generate for each instrument  (\S\ref{sec:instru}) mock data to a given S/N at its fiducial wavelength. The S/N at other wavelengths is then calculated from the SED of the individual source. We do not consider telescope or astrophysical backgrounds, or confusion noise. This is equivalent to assuming that the source flux densities can be measured to the quoted noise level by a combination of observing parameters and source extraction methods. This approach affords broad comparisons across wavelength ranges and redshifts as it is agnostic to specific telescopes or survey designs, but restricts how our results should be interpreted, as detailed below. 

Telescope and astrophysical backgrounds can limit observations, but the degree depends on the observed wavelength, the brightness of the source, and the telescope. To illustrate, we show in Figure \ref{fig:backgrounds} the SED of the ULIRG Mrk 231, together with typical astrophysical backgrounds, and telescope backgrounds for three representative facilities.  Depending on the parameters of the observation, the source can be orders of magnitude brighter than these backgrounds, or well below them.  Furthermore, backgrounds can be substantially reduced by methods such as nodding and chopping \citep{nodding83}. In not considering these backgrounds, we assume that the combination of telescope properties and observing parameters mean that these backgrounds can be removed so that their effect is insignificant. This is likely a reasonable assumption in most cases but may be invalid for very high redshift sources observed at longer wavelengths with warmer telescopes. 
 
Source confusion arises when there are secondary sources in the instrument beam, in addition to the primary source \citep{condon74}. It increases in importance as angular and/or spectral resolution become coarser and exposure time increases \citep{hogg01,takeuchi04,kiss05,bethermin17,bonatto24}.  The confusion limit used to be a straightforward bound on how deeply observations could usefully be performed, as it depended primarily on observed wavelength, and telescope aperture, \& temperature. In the last ten years, this has changed. Modern approaches combine advanced statistical methods with prior information including redshift, luminosity,  position, angular extent, and SED shape. This allows flux densities to be measured to well below the classical confusion limit \citep{hurley17,pearsonw17,pearsonw18,weaver23farm,donna24,beth24,wangl24}. Exactly how far below, however, depends on the instrument, the observations, and the source under study. The real-world equivalents of the instruments in our study vary substantially in angular resolution and sensitivity, even for the same observed wavelengths. Accounting for source confusion in our study, in a way relevant to modern analysis methods, would involve simulating a map of each observation at each wavelength with an assumed angular resolution, applying modern source extraction methods to the ensemble of observations, and then fitting the resulting SEDs. Such an approach is beyond the scope of this work and would render the comparisons across wide wavelength ranges impractical.   In not accounting for source confusion, we assume that the instrument, observation, and source properties are such that the observations can be made to the assumed S/N. This approach means our results cannot be directly used to determine the effectiveness of a survey to a fixed flux density limit - the role of source confusion must first be independently estimated to see if observations of a given S/N can be extracted.  For this reason, we do not present the luminosity limits as a function of redshift for our results. 

\begin{figure*}
\begin{center}
\includegraphics[width=0.98\linewidth]{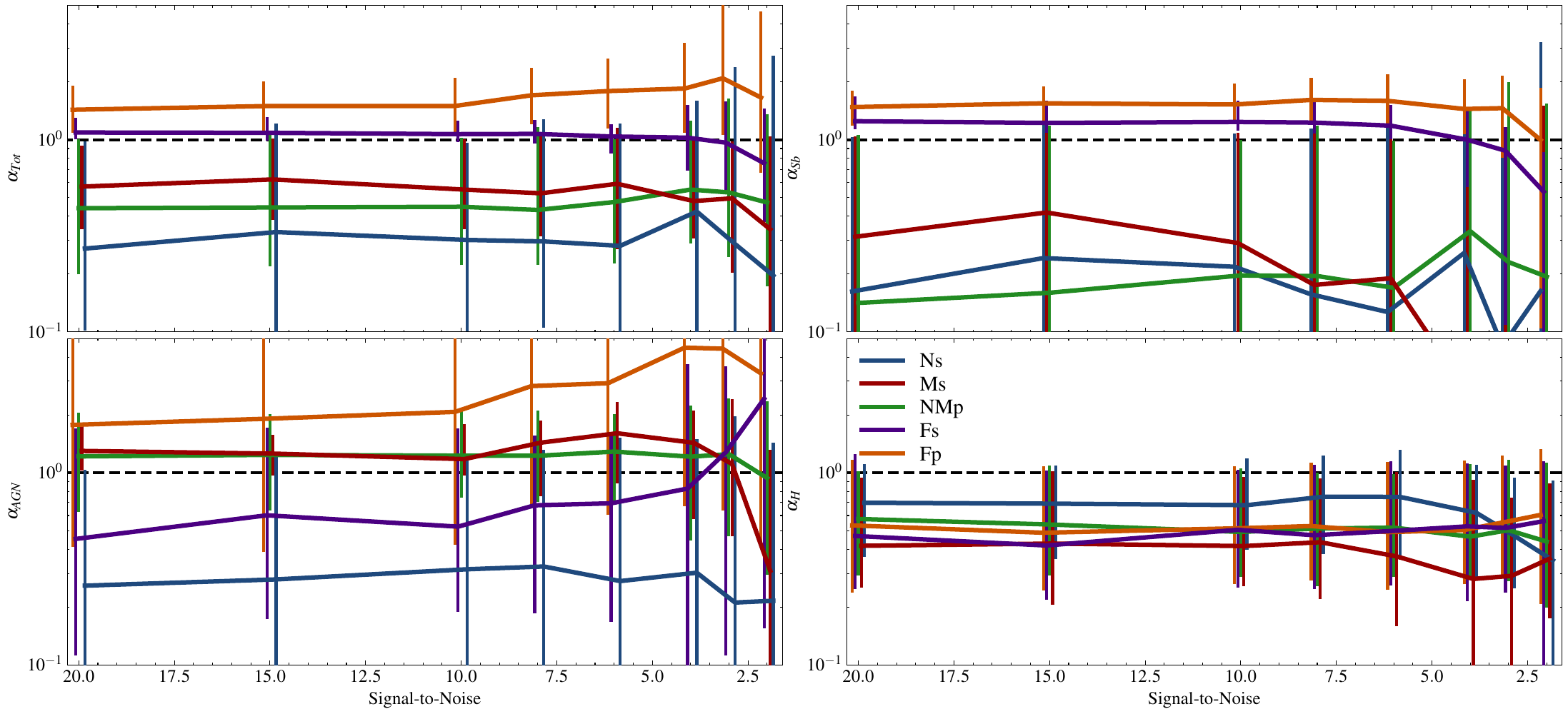}  
\caption{The effect of varying the S/N of the simulated observations on the recovery of total, starburst, AGN, and host luminosity (\S\ref{sec:ressnr}). The lines have been slightly offset in the $x$-axis for clarity.}
\label{fig:localsingle_snr_comp}
\end{center}
\end{figure*}

\begin{figure*}
\begin{center}
\includegraphics[width=0.98\linewidth]{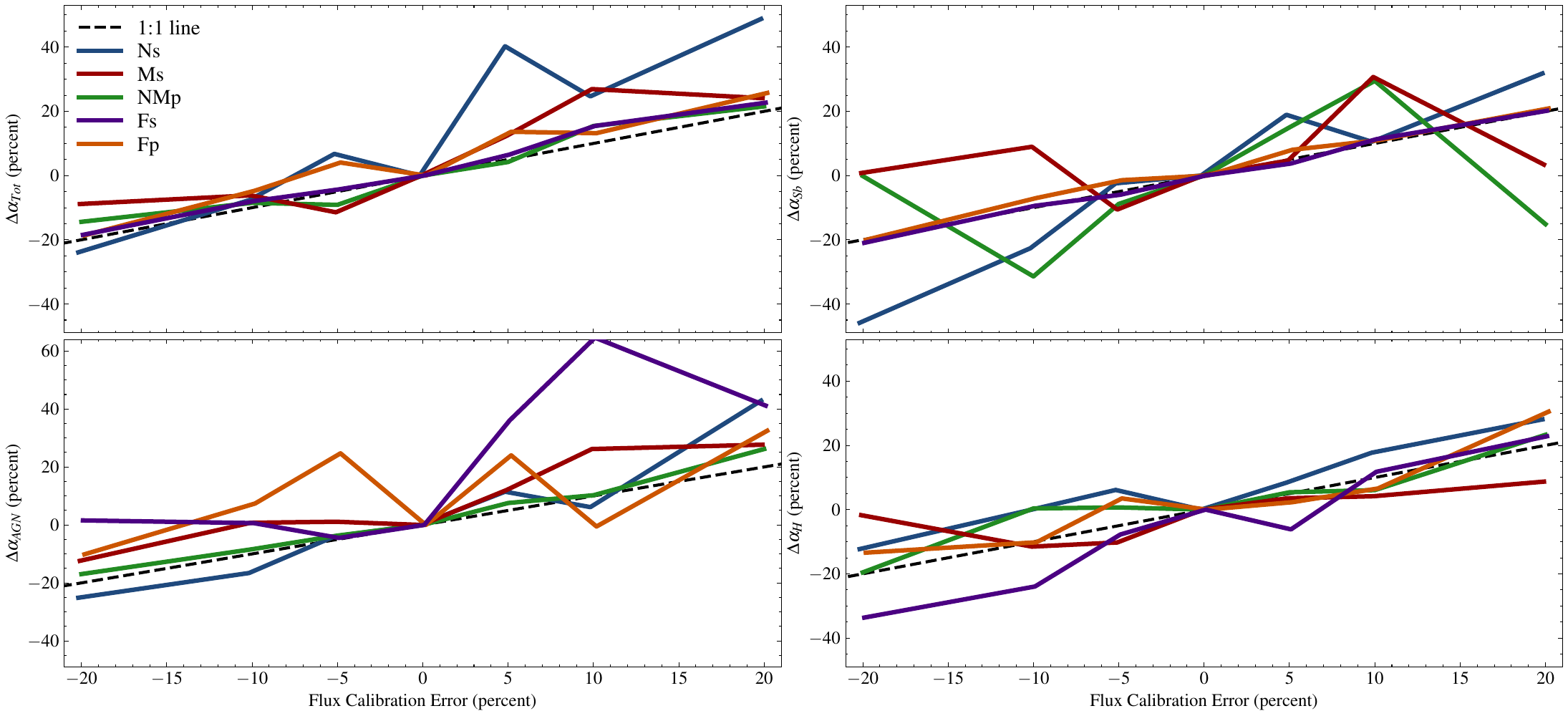}  
\caption{The effect of a flux calibration error on the recovery of total, starburst, AGN, and host luminosity (\S\ref{sec:resfcalib}). The lines have been slightly offset in the $x$-axis for clarity. Error bars have also been omitted for clarity but are comparable in size to those for the S/N=10 observations in Figure \ref{fig:localsingle_snr_comp}.}
\label{fig:localsingle_fcalib_comp}
\end{center}
\end{figure*}

\section{Parameter recovery}\label{sec:obsrec}
We fit the simulated observations with the SMART SED fitting code \citep{smart24}. SMART is an updated version of the code used in \citet{efs22}, and uses the same SED libraries.  Moreover,  \citet{varna24}  compared key physical quantities derived by SMART with those in \citet{efs22}, and found very good agreement between the estimates.  Thus, our results should reflect the degree to which different instruments can recover starburst, AGN, and host properties, rather than differences in model libraries. We restrict to the CYGNUS AGN model set with its assumed smooth tapered disk geometry since these were used to produce the `true' observations.  This means our results do not include error terms for not knowing the true AGN obscurer geometry. These uncertainties may be significant \citep[e.g.][]{gaowang21}. We defer an analysis of this issue to a dedicated future work. An example `true' SED, the simulated data, and the recovered SED fit to these simulated data are shown in  Figure \ref{fig:examplesed}.

To assess the quality of the recovery, we compare the following ``observed" luminosities to their ``true" values:

\begin{itemize}

\item $L_{Tot}$: The total rest-frame $1-1000\mu$m luminosity of the system. 
	
\item $L_{Sb}$: The rest-frame $1-1000\mu$m luminosity of the starburst.

\item $L_{AGN}$: The rest-frame $1-1000\mu$m luminosity of the AGN. This measure of AGN luminosity does not include a correction for anisotropic emission.  Recovering the anisotropy-corrected AGN luminosity requires the observations to constrain the geometry of the AGN obscurer. We do not consider this here, but defer it to a future work. 

\item $L_{H}$: The rest-frame $1-1000\mu$m luminosity of the host galaxy. 

\end{itemize}

\noindent The other parameters for each model are allowed to vary in the fits, using the same ranges as in \citet{efs22}, but we focus here on how well total and component infrared luminosities can be recovered.

\begin{figure*}
\begin{center}
\includegraphics[width=0.98\linewidth]{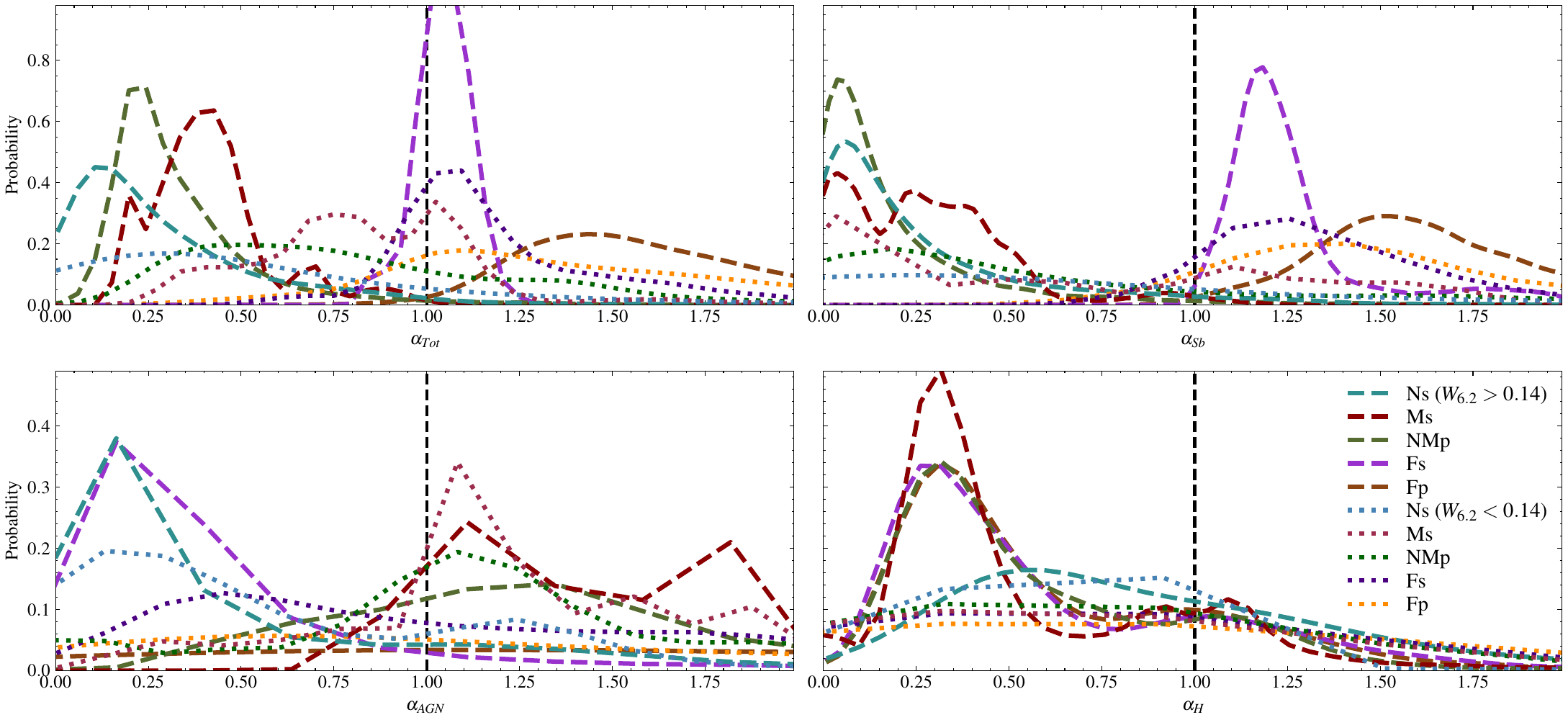} 
\caption{As Figure \ref{fig:localsingle} but using two subsets of the full input population - one with $6.2\mu$m PAH EWs greater than $0.14\mu$m (dashed), and one with $6.2\mu$m PAH EWs less than $0.14\mu$m (dotted). Axis ranges have been modified to aid comparison (\S\ref{sec:paheffect}).}
\label{fig:localsingle_pah}
\end{center}
\end{figure*}

\begin{figure*}
\begin{center}
\includegraphics[width=0.49\linewidth]{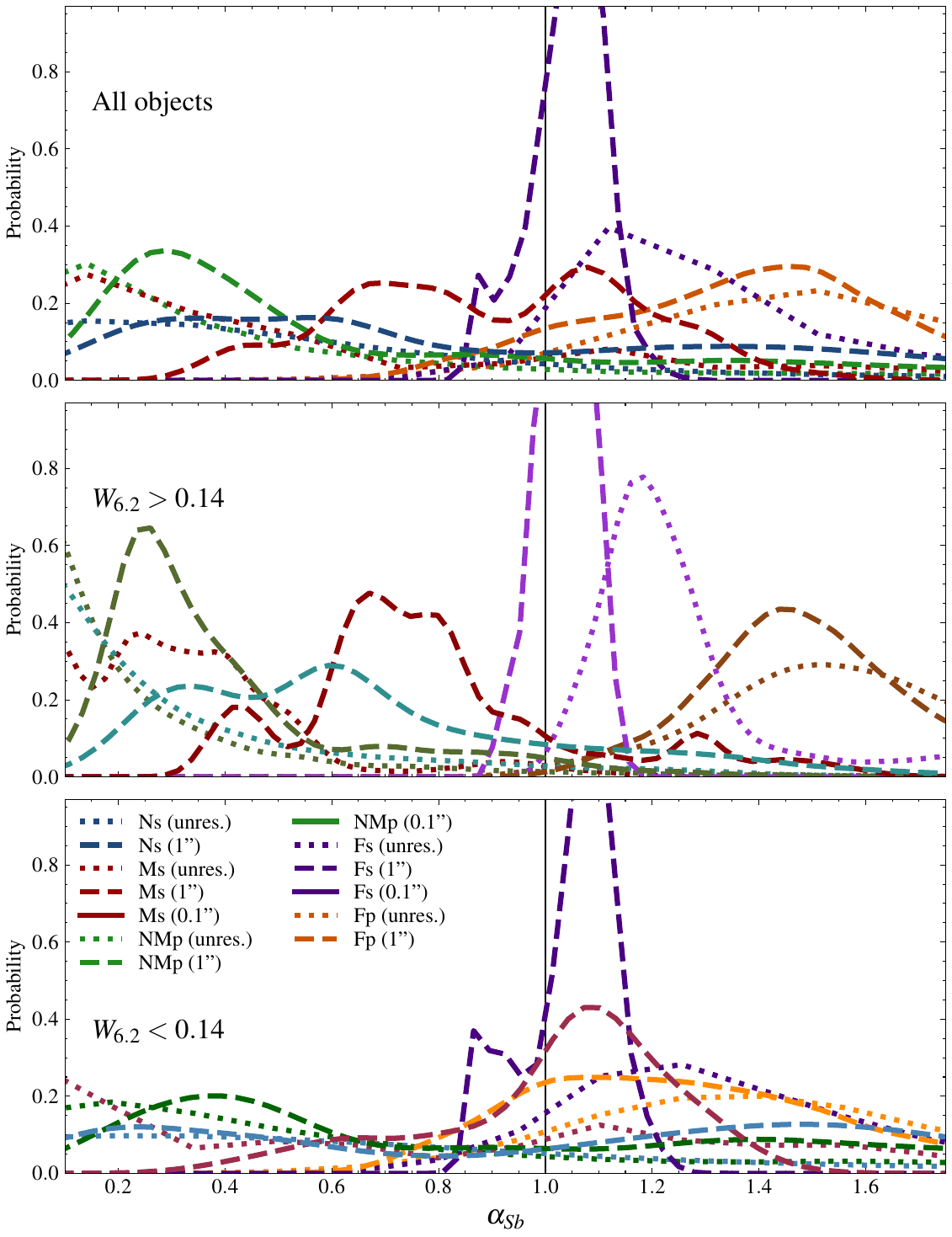} 
\includegraphics[width=0.49\linewidth]{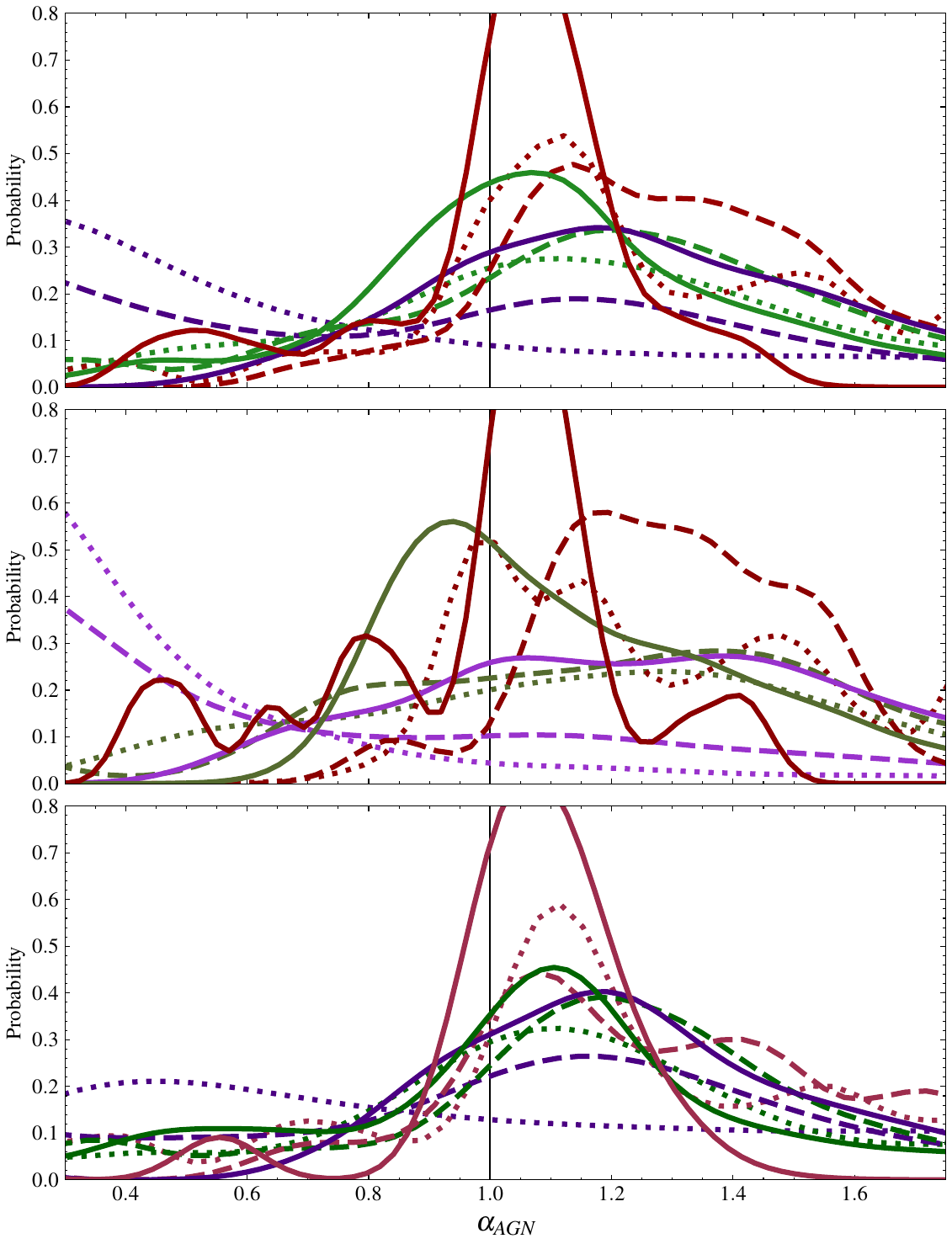} 
\caption{The effect on the recovery of (left column) starburst and (right column) AGN luminosity of observing at $1\arcsec$ angular resolution for $L_{Sb}$, and $1\arcsec$ and $0.1\arcsec$ resolution for $L_{AGN}$ (\S\ref{sec:resspatres}). The three rows show results for simulations using all objects, and the high PAH EW and low PAH EW subsamples.  The results for Ns and Fp are not plotted as the finer angular resolution does not change their results significantly.}
\label{fig:localsingle_spatial}
\end{center}
\end{figure*}

\section{Results}\label{sec:res}

\subsection{Nearby Universe}\label{sec:resnear}
We first examine the ability of observations with any one of Ns, NMp, Ms, Fs, and Fp (plus Opt in all cases) to recover total infrared luminosities in the nearby Universe ($z\sim0.1$). We parametrize this recovery as the ratio of the observed to true luminosity:

\begin{equation}\label{eq:alphadef}
\alpha = \frac{L^{Observed}}{L^{True}}
\end{equation}

\noindent We assume that each observation is detected at an S/N of $10\sigma$ at its fiducial wavelength.  The results are shown in Figure \ref{fig:localsingle} and Table \ref{tbl:highzsingle}. The Fs instrument gives the best recovery, constraining $L_{Tot}$ to within 10\% and with little systematic bias. In contrast, the recovery of $L_{Tot}$ with Fp is biased high by about 50\%, and low by about 50\% with NMp or Ms, on average, with random uncertainty of $\sim30-40\%$. With Ns the recovery of $L_{Tot}$ is biased low by about $70\%$.  For $L_{Sb}$,  Fs and Fp overestimate by 25\% and 50\%, on average, Ms and NMp underestimate by 50\% and 70\%, and Ns gives no useful constraints.   For $L_{AGN}$, Ms and NMp overestimate by $20-30\%$, Fs and Ns underestimate by about $30\%$, and Fp gives no useful constraints. For $L_{H}$, Ns gives the most accurate recovery, followed by Fs, Fp, NMp, and Ms. All observations, though, systematically underestimate $L_{H}$, by at least 20\%. 

\subsubsection{Effect of Signal-to-Noise}\label{sec:ressnr}
We next explore the effect of varying the S/N of the simulated observations over the range $2-20$ (Figure  \ref{fig:localsingle_snr_comp}). Increasing the S/N from 10 to 20 has no significant impact. Reducing the S/N much below 10, however, does. For all instruments, there is noticeable degradation in the recovery of total, starburst, AGN, and host luminosity by S/N$\simeq6$, and substantial degradation by S/N$\simeq4$. Reducing the S/N impacts precision (random error) more than accuracy (systematic bias). Since we do not want our subsequent results to be dominated by S/N effects, we fix S/N=10 for all simulations in the nearby universe.

\subsubsection{Effect of Flux Calibration}\label{sec:resfcalib}
Another potential issue is an error in the flux calibration of the instrument. To investigate this, we explore the impact on luminosity recovery of a uniform flux calibration error of between $-20$\% and $20$\% over the wavelength range of each instrument (Figure \ref{fig:localsingle_fcalib_comp}).  For $L_{Tot}$, fitting a linear model between the flux calibration error and the bias in luminosity recovery yields a slope consistent with unity to within $1\sigma$ for Ms, NMp, Fs, and Fp, and to within $2\sigma$ for Ns.  For $L_{Sb}$ the slopes are consistent with unity to within $1\sigma$ for Fs and Fp, but show tension with a unity slope at the $2-2.5\sigma$ level for Ns, Ms, and NMp.  For $L_{AGN}$ the slopes are consistent with unity to within $1\sigma$ for all instruments except Ns, which is consistent at $\sim2\sigma$. For $L_{H}$ the slopes are consistent with unity to within $2\sigma$ for all instruments except Ms, which is inconsistent with unity at the $\sim3.5\sigma$ level. These results may suggest the presence of a subtle effect, in which a flux calibration error in a given instrument more strongly impacts the recovery of SED luminosities that are faint in its wavelength range than those that are bright. A study with a larger sample would, however, be required to confirm this.

\begin{figure*}
\begin{center}
\includegraphics[width=0.49\linewidth]{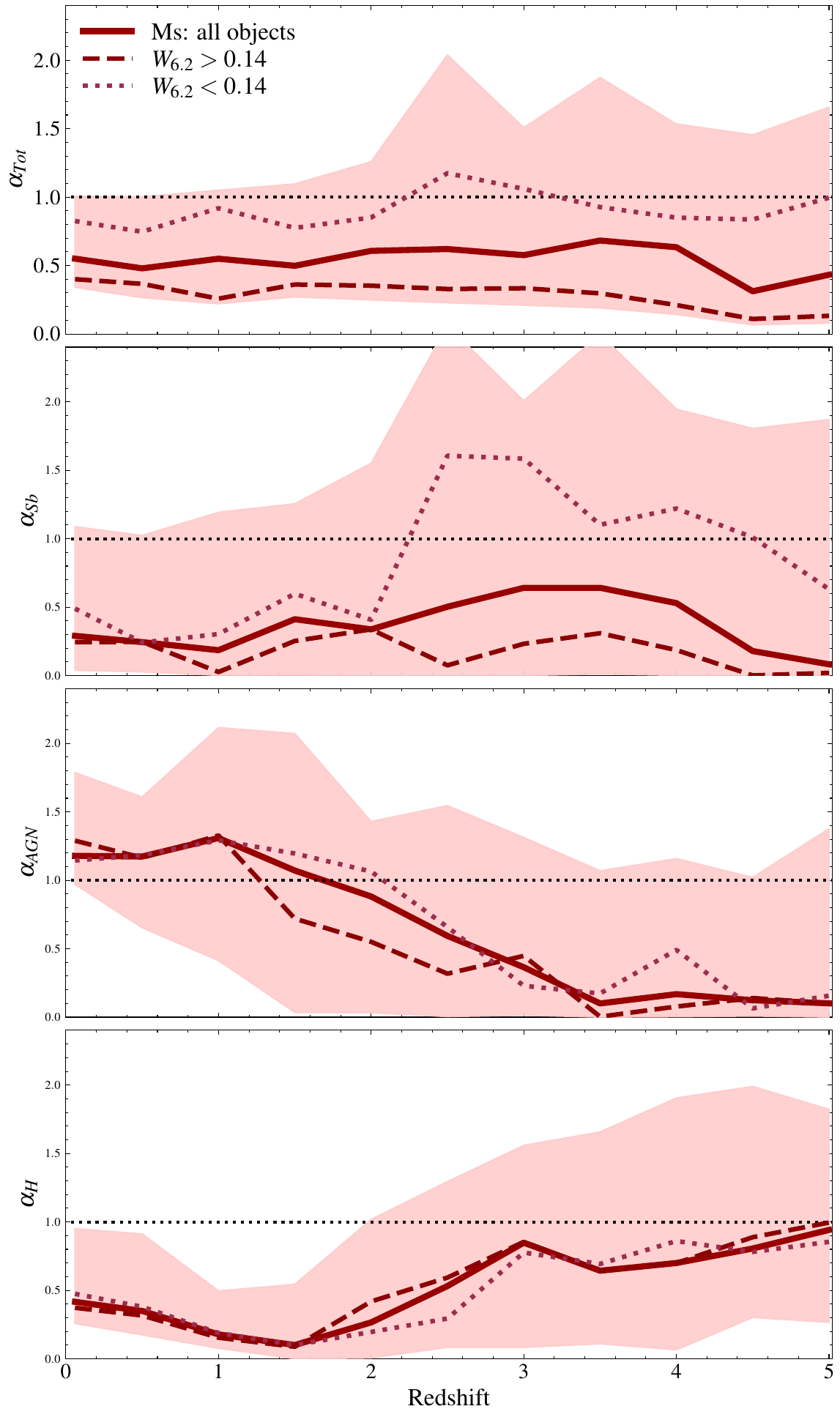} 
\includegraphics[width=0.49\linewidth]{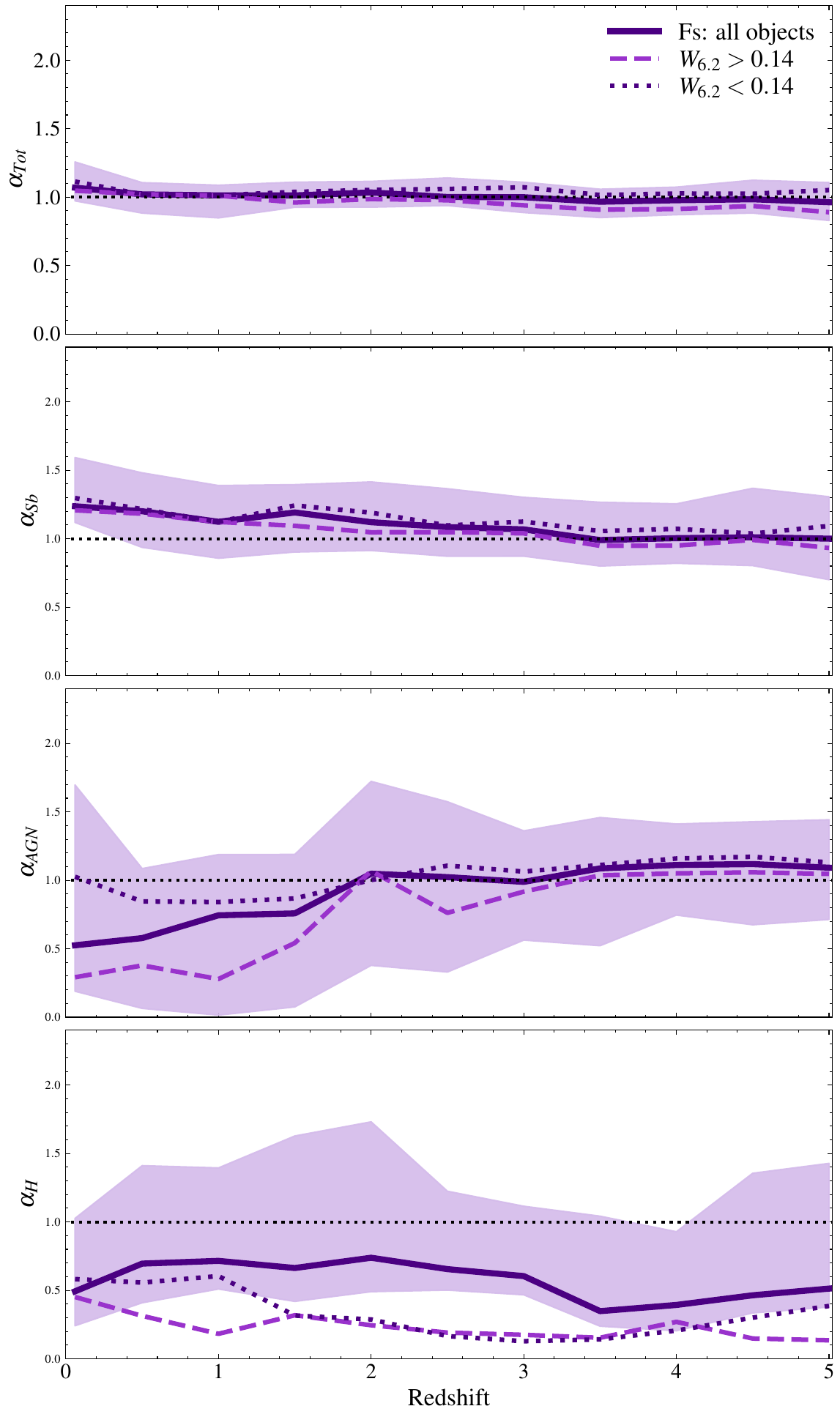} 
\caption{The recovery of total, starburst, AGN, and host galaxy luminosity using (left column) Ms, and (right column) Fs, over $0<z<5$ (\S\ref{sec:resfar}). The solid line and shaded region show the results from simulations using all objects and the $1\sigma$ error region, The dashed and dotted lines show the results for the high and low PAH EW sub-samples.}
\label{fig:redshiftsingle_spects}
\end{center}
\end{figure*}

\subsubsection{Effect of Starburst vs. AGN dominance}\label{sec:paheffect}
The initial selection of an object may predispose it towards starburst or AGN dominance. Simulations with an input population that spans the full range of starburst to AGN dominance may not be applicable in these cases.  Accordingly, we divide our input sample into two sub-samples; one with $6.2\mu$m PAH equivalent widths (EWs) of $W_{6.2} > 0.14\mu$m and one with $W_{6.2} < 0.14\mu$m. We adopt this criterion as PAH EW is commonly used to select for starburst dominance \citep{desai07,bern09,takagi10,leiceers24}, and because this boundary closely divides the sample into systems where more (less) than half of $L_{Tot}$ arises from star formation.  This does however mean that the results from each simulation are based on only 21 objects. This should still be sufficient to estimate the recovery of luminosities, but we note this as a caveat to our analysis. 

The results are shown in Figure \ref{fig:localsingle_pah} and Table \ref{tbl:highzsingle}. For the $W_{6.2} > 0.14\mu$m simulations, the recovery of $L_{Tot}$ and $L_{Sb}$ is biased towards lower values using Ns, Ms, or NMp, but effectively unchanged for Fs and Fp. For $L_{AGN}$, Ns and Fs become more biased to lower values, while Ms and NMp are unchanged.  The recovery of $L_{H}$ is unchanged, for any instrument.  For the $W_{6.2} < 0.14\mu$m simulations, the results are different. For $L_{Tot}$ and $L_{Sb}$, the Ns, Ms, and NMp instruments all give less biased recovery, while the recovery using Fs and Fp is barely changed.  For $L_{AGN}$, the recovery with Ns and Fs is slightly less biased, but unchanged with Ms, NMp, or Fp. The recovery of $L_{H}$ is not affected, using any instrument.  

\begin{figure*}
\begin{center}
\includegraphics[width=0.49\linewidth]{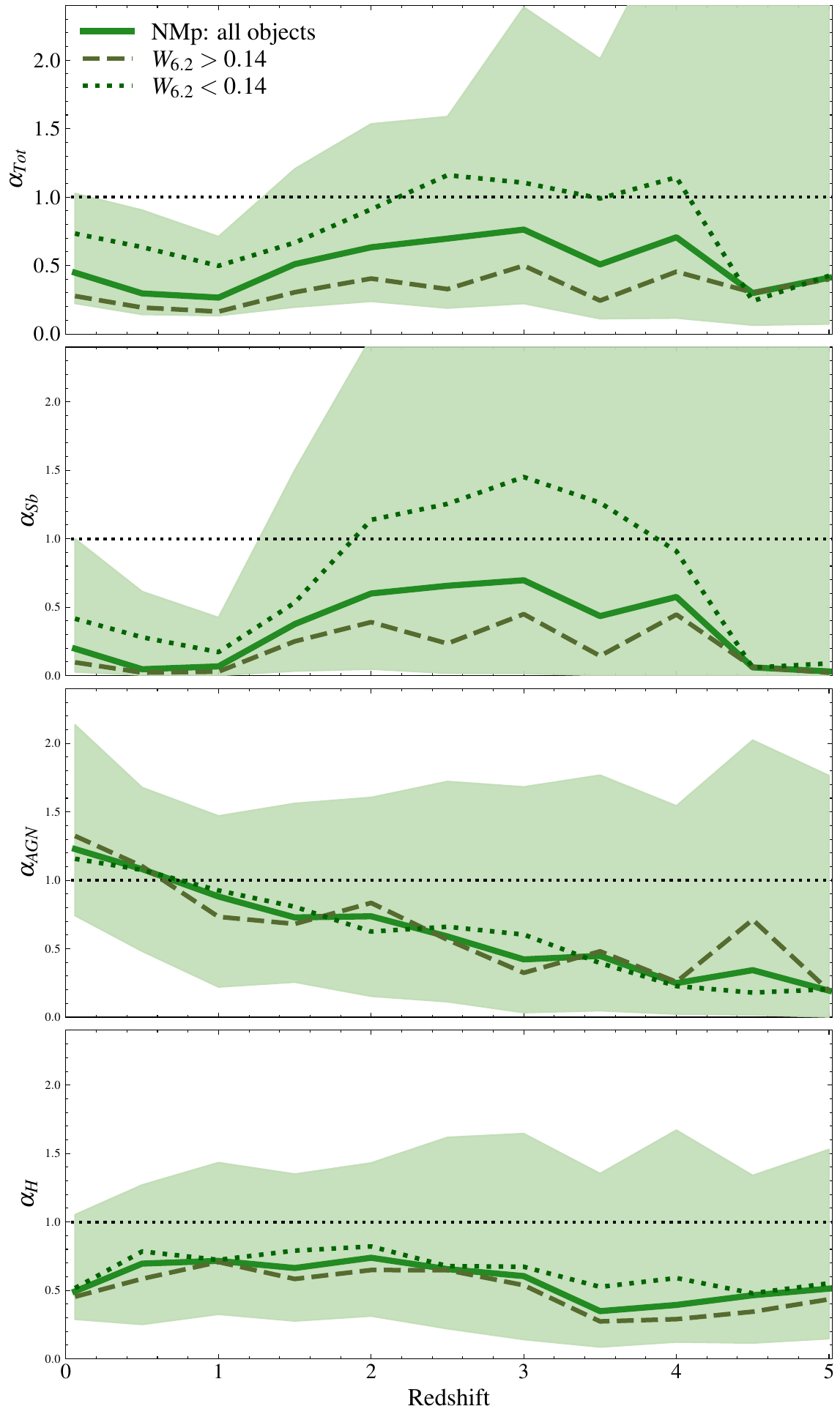} 
\includegraphics[width=0.49\linewidth]{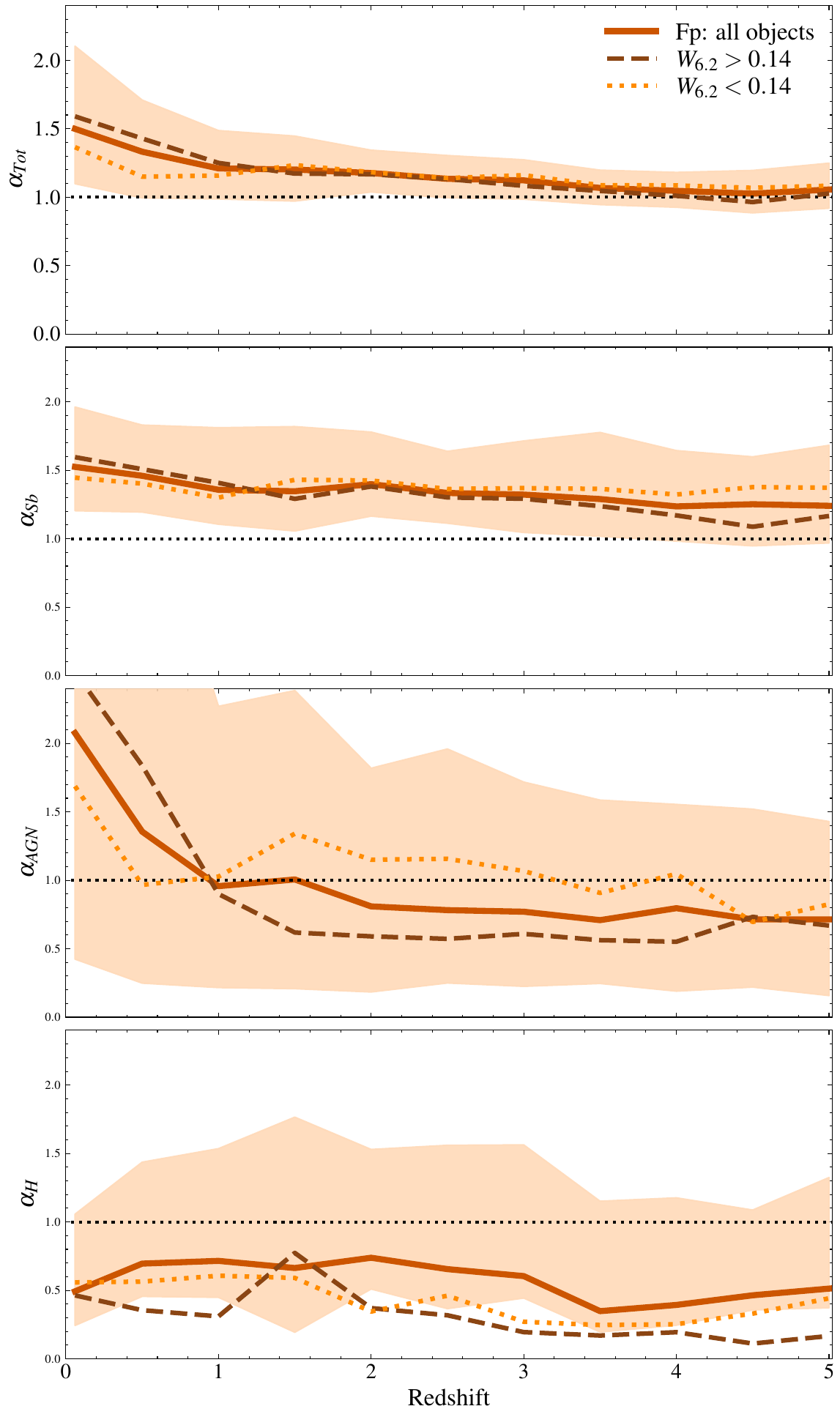} 
\caption{As Figure \ref{fig:redshiftsingle_spects} but for the (left column) NMp and (right column) Fp instruments.}
\label{fig:redshiftsingle_phots}
\end{center}
\end{figure*}

\begin{figure}
\begin{center}
\includegraphics[width=0.98\linewidth]{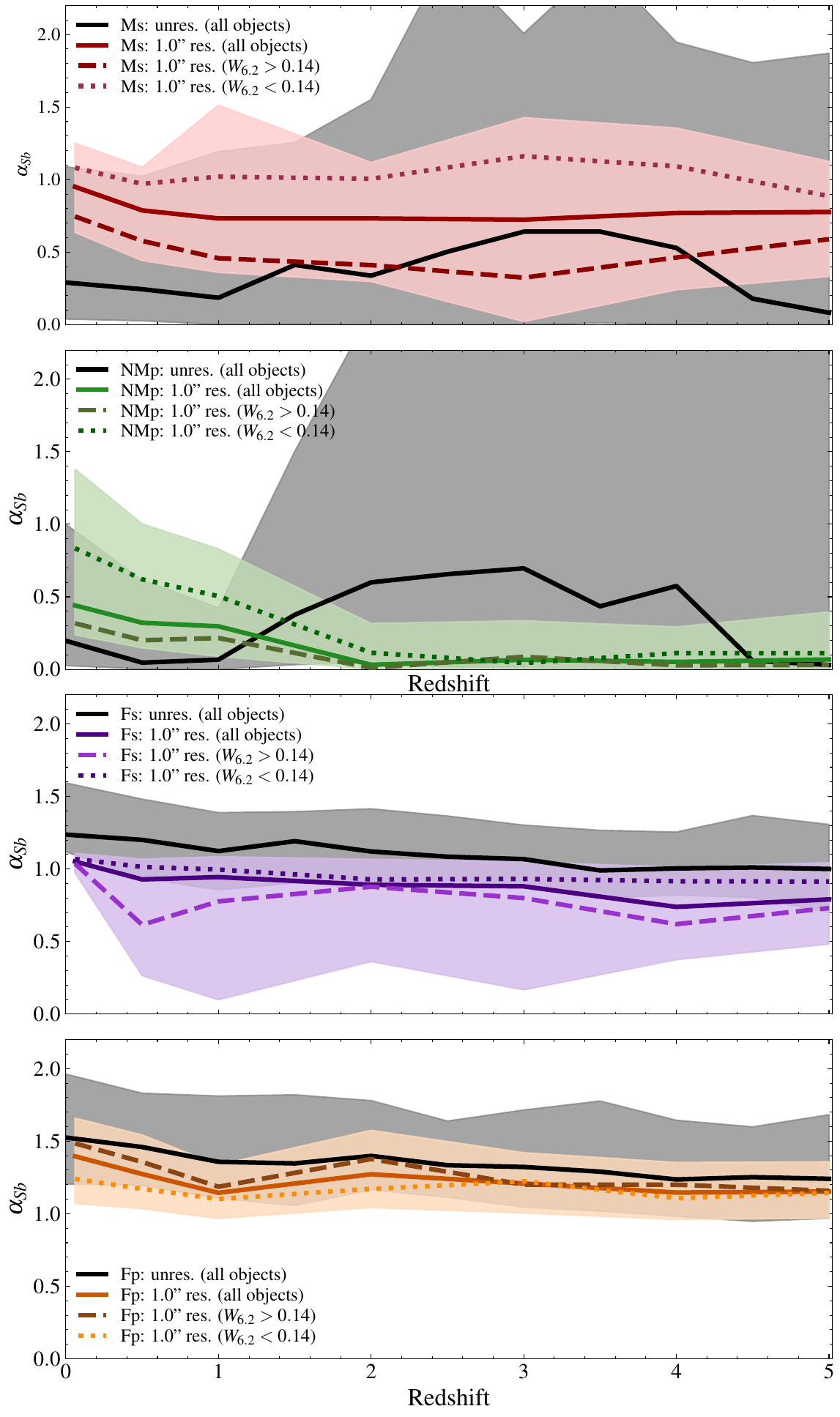} 
\caption{The effect of angular resolution on the recovery of starburst luminosity as a function of redshift, using the Ms, NMp, Fs, and Fp instruments (\S\ref{sec:resfar}). Each panel compares the  spatially unresolved results using all objects for starburst luminosity from Figure \ref{fig:redshiftsingle_spects} with those obtained from observing at $1.0\arcsec$ resolution using all objects, and the high and low PAH EW sub-samples.}
\label{fig:zsing_spat1}
\end{center}
\end{figure}

\subsubsection{Effect of Angular Resolution}\label{sec:resspatres}
We here examine the effect on luminosity recovery of observing at finer angular resolution. We consider the scenario where finer angular resolution allows a component SED to contribute more of the emission in an aperture. We focus on the recovery of $L_{Sb}$ and $L_{AGN}$ under the assumption that the host is much larger than the starburst or AGN.  We adopt a simple model in which each galaxy has a host of  diameter $10$\,kpc, a starburst of diameter $1$\,kpc, and an AGN of diameter $0.1$\,kpc.  We assume the light from each component is uniformly distributed over these regions. For ease of comparison with the other results in \S\ref{sec:res} we keep each galaxy at its redshift in Table \ref{tbl:sample}, though moving them all to a common redshift of $z=0.1$ has an insignificant effect on the results that follow. Then, for $L_{Sb}$, we consider an angular resolution of $1\arcsec$, centered on a starburst + host region  (finer angular resolutions start to resolve the starburst). For $L_{AGN}$ we consider angular resolutions of $1\arcsec$ and $0.1\arcsec$, centered on a starburst+AGN+host region. 
 
The results are shown in Figure \ref{fig:localsingle_spatial} and Table \ref{tbl:highzsingle}. For $L_{Sb}$, the $1\arcsec$ resolution reduces systematic bias for all instruments.  The effect is most marked for the spectrometers, especially for Ms, for which the systematic bias becomes insignificant. The high PAH EW simulations predict generally larger systematic biases, and the low PAH EW simulations predict smaller systematic biases.  For $L_{AGN}$, the recovery with Ns is only marginally improved by observing at $1.0\arcsec$ or $0.1\arcsec$ resolution. In contrast, with Ms and NMp the recovery is noticeably  improved at $1\arcsec$ resolution, and improved further at $0.1\arcsec$ resolution. With Fs, there is slight improvement at $1\arcsec$ resolution and significant improvement at $0.1\arcsec$ resolution.   With Fp, the improvement is marginal.  We again see a tendency for the high PAH EW simulations to show more biased recovery, and the low PAH EW simulations to show less biased recovery, though the effect is small at $0.1\arcsec$ resolution.  An important corollary is that, since we are not including source confusion, these results are lower bounds on the improvement that finer angular resolution provides.

\begin{figure}
\begin{center}
\includegraphics[width=0.98\linewidth]{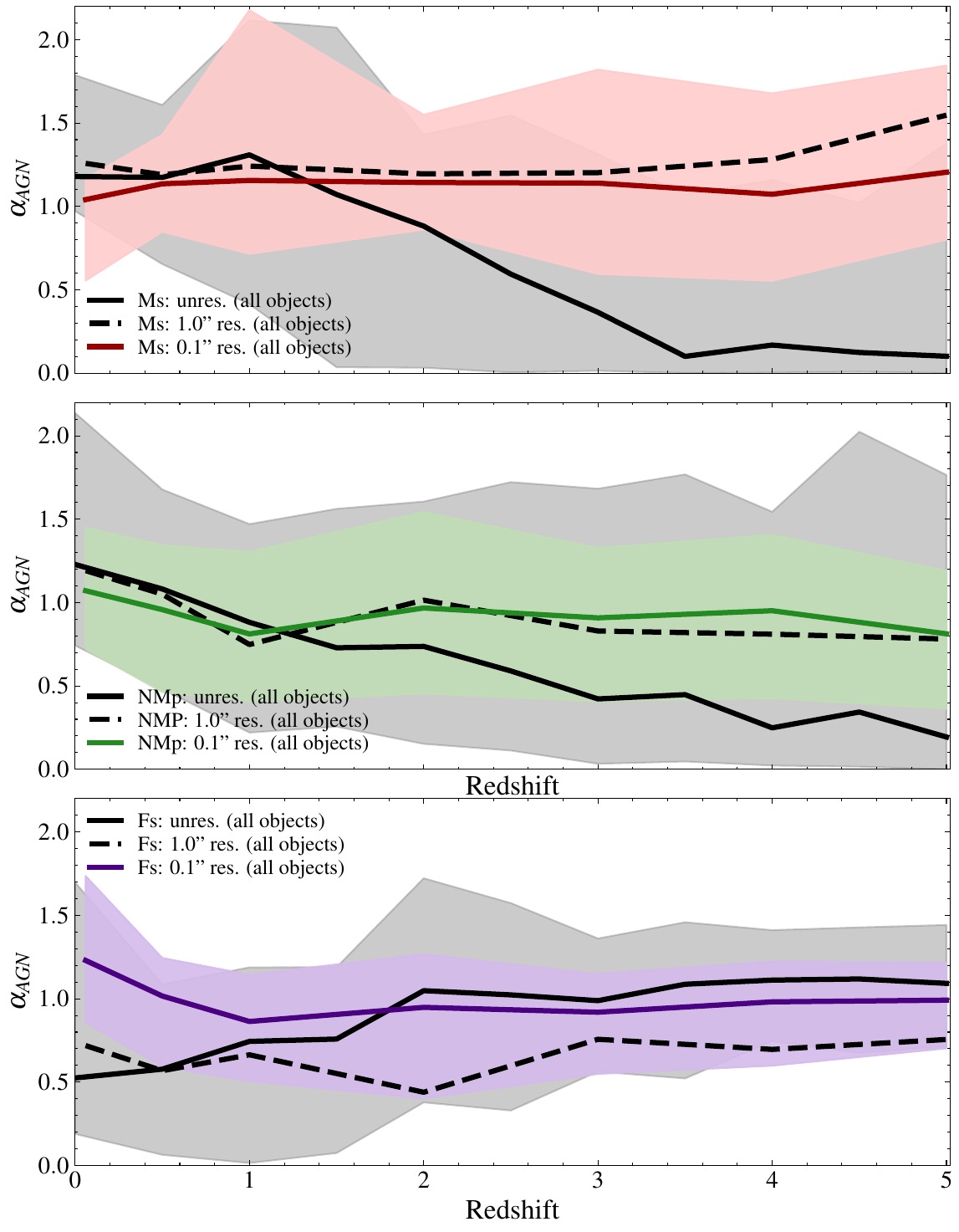} 
\caption{The effect of angular resolution on the recovery of AGN luminosity as a function of redshift, using the Ms, NMp, and Fs instruments (\S\ref{sec:resfar}). Each panel compares the spatially unresolved results using all objects for AGN luminosity from Figure \ref{fig:redshiftsingle_spects} with those obtained from observing at $1.0\arcsec$ and $0.1\arcsec$ resolution using all objects (the results for the high and low PAH EW sub-samples are omitted as they lie close to their respective results using all objects).}
\label{fig:zsing_spat2}
\end{center}
\end{figure}

\begin{figure*}
\begin{center}
\includegraphics[width=0.98\linewidth]{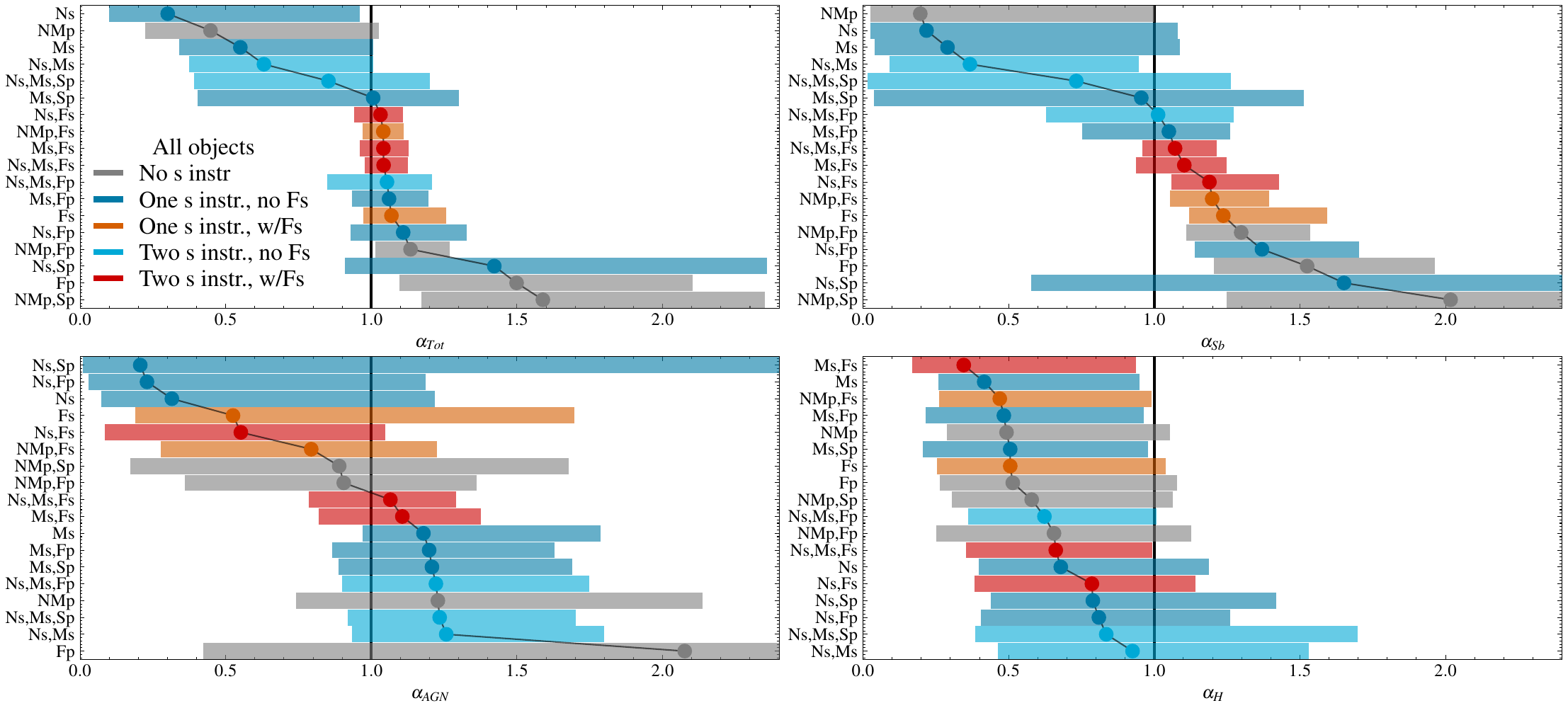} 
\caption{The recovery of total, starburst, AGN, and host luminosity at $z\sim0.1$, using the combinations of instruments specified on the y-axis (\S\ref{sec:multi}). In each panel, the results are ordered by their $\alpha$ value. The error bars are the $1\sigma$ uncertainties.}
\label{fig:localmulti_all}
\end{center}
\end{figure*}

\subsection{High-redshift Universe}\label{sec:resfar}
We here examine luminosity recovery using single-instrument observations over $0 < z < 5$. For these simulations, we assume  $6\sigma$ detections at the observed-frame fiducial wavelengths. This is a compromise to the dimming effects of redshift without entering a regime where the reduced S/N dominates the results. We do not consider the Ns instrument as we found it provides no meaningful constraints on luminosities at $z\gg0.1$. 

The results are shown in Figures \ref{fig:redshiftsingle_spects} and  \ref{fig:redshiftsingle_phots}, and  in Table \ref{tbl:highzsingle}. For $L_{Tot}$, Ms and NMp gives luminosities that are biased low by about 50\% at all redshifts. The high (low) PAH EW simulations predict more (less) bias, respectively.  In contrast, Fs gives measures of $L_{Tot}$ that are virtually unbiased at all redshifts, and shows insignificant difference between the high and low PAH EW sub-samples. The Fp instrument gives $L_{Tot}$ estimates that are biased high at lower redshifts, but this bias reduces to nearly zero by $z\sim5$. For $L_{Sb}$, we obtain similar results as for $L_{Tot}$: Ms and NMp give luminosities that are biased low by about 50\% at most redshifts, Fs gives almost unbiased results, and Fp is biased high at most redshifts.  For $L_{AGN}$, Ms and NMp observations at low redshift  have small systematic bias, but the bias becomes more negative as redshift increases.  In contrast, the $L_{AGN}$ recovery with Fs is systematically biased low at low redshifts, but this bias diminishes as redshift increases.  In all three cases, the low PAH EW simulations again show less systematic bias than the high PAH EW simulations.  Observations with Fp give $L_{AGN}$ values that are systematically biased low at all redshifts, and with large random errors. 

We next examine how finer angular resolution affects the recovery of $L_{Sb}$ and $L_{AGN}$ as a function of redshift.  The results are shown in Figures \ref{fig:zsing_spat1} and \ref{fig:zsing_spat2}, and in Table \ref{tbl:highzsingle}. Finer angular resolution improves the recovery of $L_{Sb}$ and $L_{AGN}$ for all instruments at all redshifts. For $L_{Sb}$, arcsecond resolution  significantly reduces, though does not eliminate, the systematic bias seen for Ms and Fp. There is only a small impact on the recovery with Fs or NMp. For $L_{AGN}$, there is noticeable improvement in recovery at $1\arcsec$ resolution for Ms, NMp, and Fs, and substantial improvement at $0.1\arcsec$ resolution.  We do not obtain useful constraints on $L_{AGN}$ with Fp at any angular resolution. As with \S\ref{sec:resspatres}, these results are lower bounds on the improvement from observing at finer angular resolution, due to our neglection of source confusion. 

\begin{figure*}
\begin{center}
\includegraphics[width=0.49\linewidth]{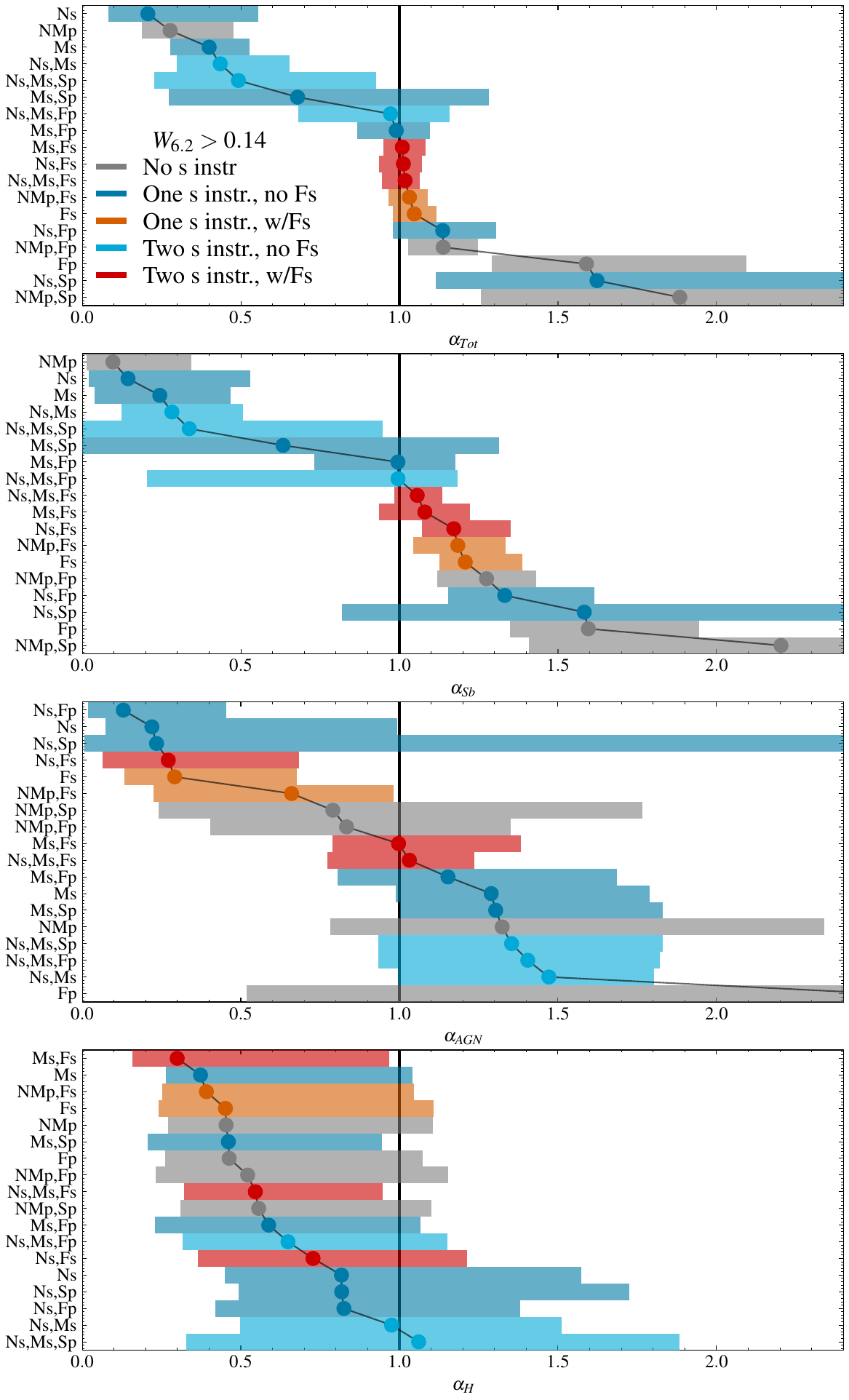} 
\includegraphics[width=0.49\linewidth]{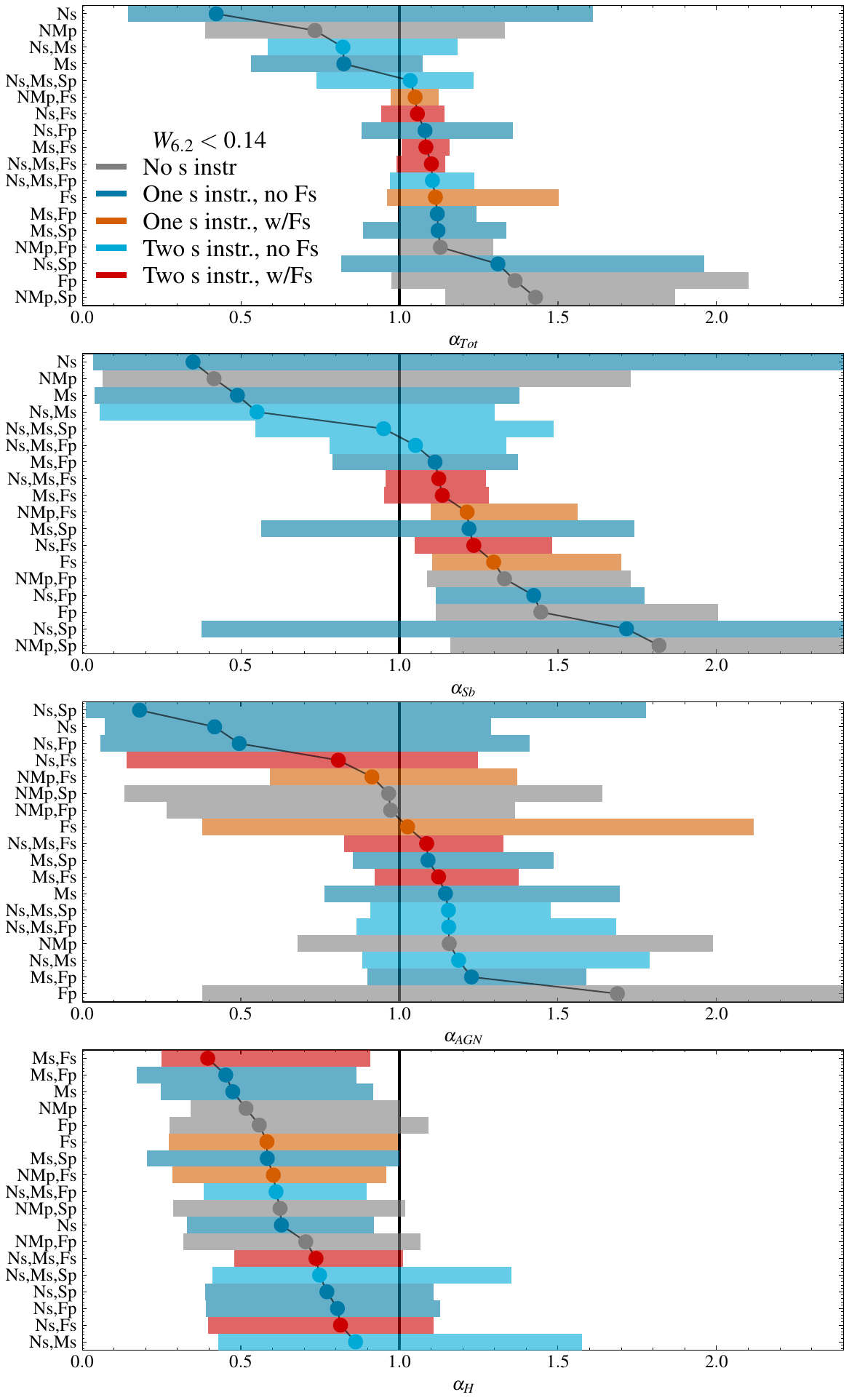} 
\caption{As Figure \ref{fig:localmulti_all} but for the (top rows) high PAH EW and (bottom rows) low PAH EW sub-samples.}
\label{fig:localmulti_all_hilo}
\end{center}
\end{figure*}

\begin{figure*}
\begin{center}
\includegraphics[width=0.98\linewidth]{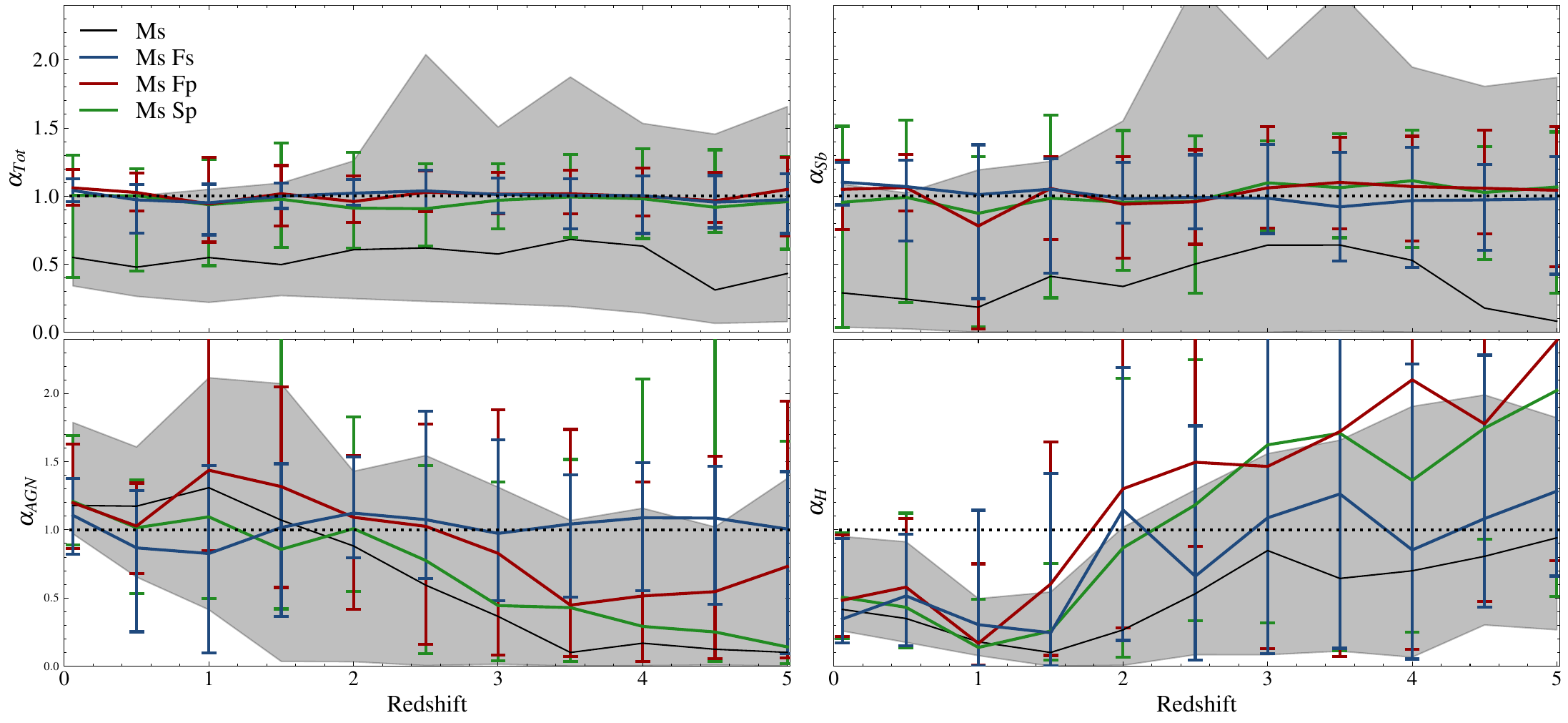} 
\caption{The recovery of total, starburst, AGN, and host luminosity as a function of redshift, using combinations of instruments where one of those instruments is Ms (\S\ref{sec:multi}). The recovery using Ms alone is included for comparison. }
\label{fig:redshiftmulti_1}
\end{center}
\end{figure*}

\begin{figure*}
\begin{center}
\includegraphics[width=0.98\linewidth]{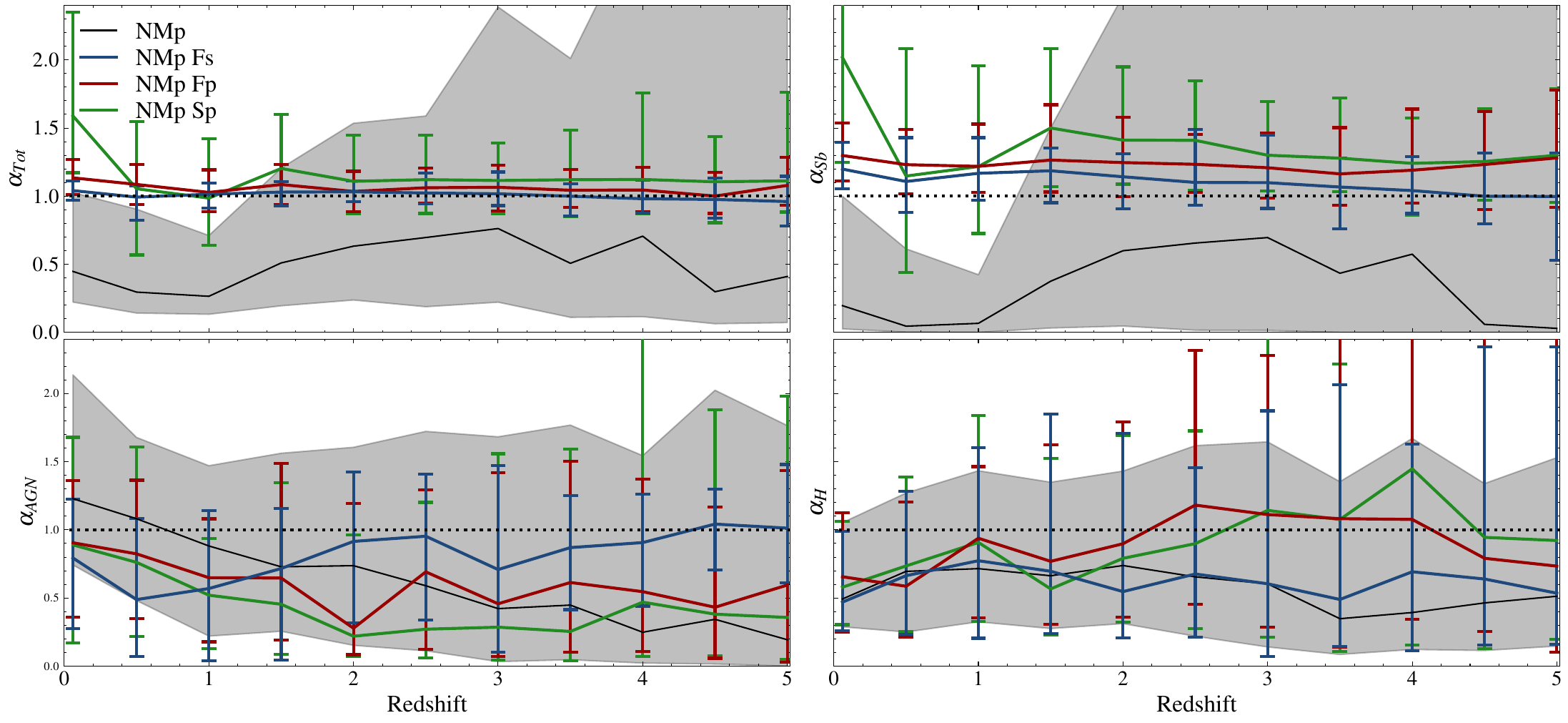} 
\caption{The recovery of total, starburst, AGN, and host luminosity as a function of redshift, using combinations of instruments where one of those instruments is NMp (\S\ref{sec:multi}). The recovery using NMp alone is included for comparison. }
\label{fig:redshiftmulti_2}
\end{center}
\end{figure*}

\subsection{Observations with multiple instruments}\label{sec:multi}

\subsection{Nearby Universe}\label{sec:resmulloc}
We here examine luminosity recovery using combinations of the instruments in Table \ref{tbl:instparams}.  We start at $z\sim0.1$, assume detections at $10\sigma$ significance at the relevant fiducial wavelength, and include observations with Opt in all cases. To keep the simulations computationally reasonable, we do not consider the effects of observing at finer angular resolution. 

The results are presented in Figure  \ref{fig:localmulti_all} and Table  \ref{tbl:localmulti}.  The best recovery of luminosities is achieved with instruments that have a long wavelength baseline, and/or dense wavelength sampling. For $L_{Tot}$, the best recovery is achieved with instrument combinations that include Fs or Fp, especially Fs. Using Fs or Fp gives much better recovery than using Sp, with the same shorter-wavelength instruments. Combinations with only photometric far-infrared observations overestimate $L_{Tot}$, while combinations with no far-infrared constraints underestimate $L_{Tot}$.  We find similar results for the recovery of $L_{Sb}$. For $L_{AGN}$, the best results are obtained for combinations that include Ns and Ms, or NMp, plus one of Fs or Fp. Combinations without a far-infrared constraint tend to bias high, whereas combinations with poor mid-infrared constraints tend to bias low.  For $L_{H}$, the best recovery is obtained with the inclusion of Ns and/or Fs, though in all cases the recovered host luminosities are biased low. While there are some minor variations, we obtain similar results when considering the high or low PAH EW simulations (Figure \ref{fig:localmulti_all_hilo}).

\subsection{High-redshift Universe}\label{sec:resmulzev}
Next, we examine the recovery of total and component luminosities using combinations of instruments over $0<z<5$.  We consider two groups; one in which Ms is paired with longer wavelength instruments (Figure \ref{fig:redshiftmulti_1}), and one where NMp is paired with longer-wavelength instruments (Figure \ref{fig:redshiftmulti_2}).  We assume all observations have an SN of 6 at their fiducial wavelengths.  First considering the Ms group: the positive effects of wavelength sampling and coverage are evident.  The addition of any of Fs, Fp, or Sp enhances the recovery of luminosities, compared to Ms alone.  However, the greatest enhancement is given by Fs, even though its wavelength baseline is shorter than Fp or Sp. Ms+Fs gives almost unbiased recovery of $L_{Tot}$, $L_{Sb}$, and $L_{AGN}$ over most redshifts. Also notable is that Ms+Fp  outperforms Ms+Sp. Turning to the NMp group, we obtain similar results, though with larger random errors. 

\begin{figure}
\begin{center}
\includegraphics[width=0.98\linewidth]{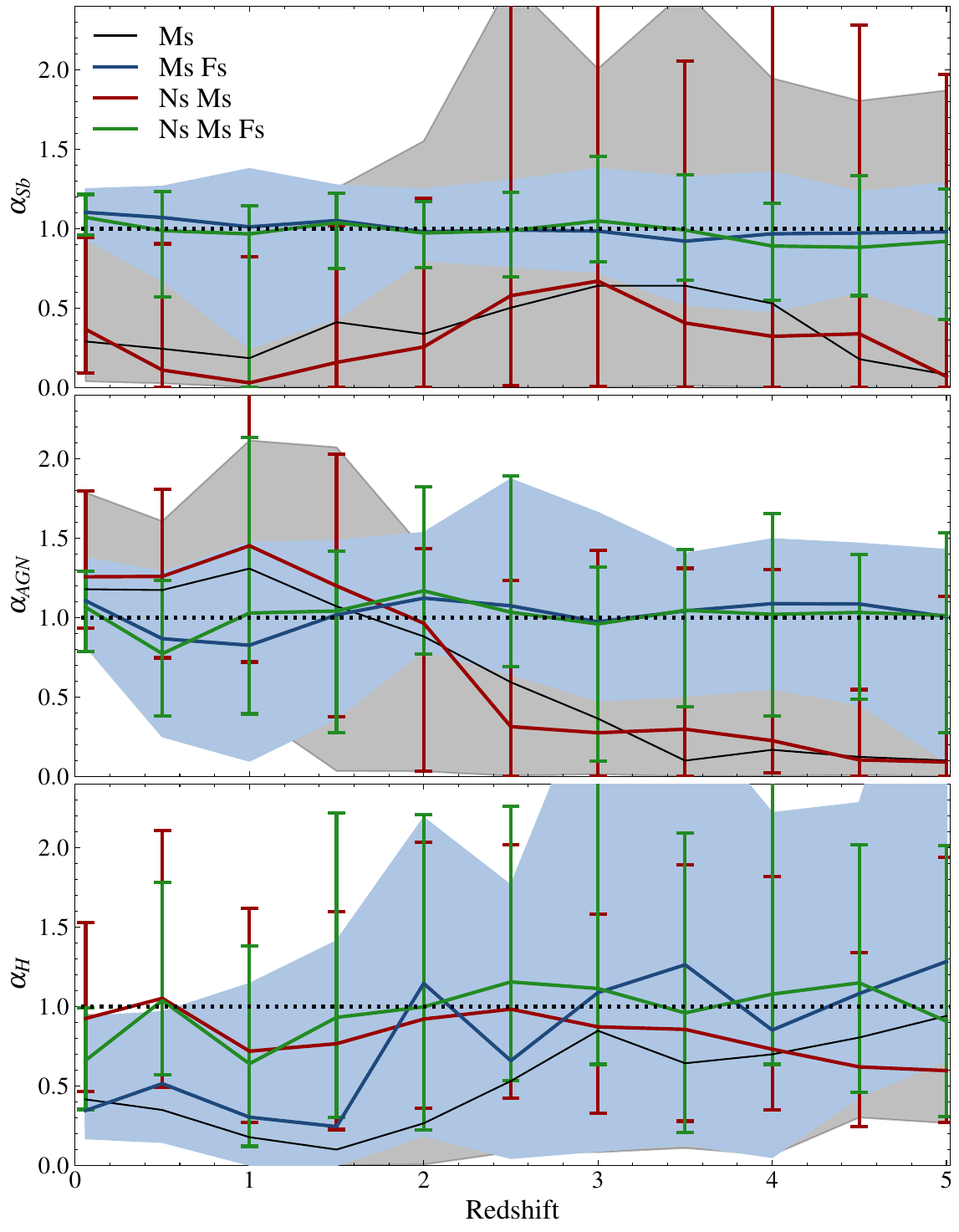} 
\caption{The effect of including Ns observations on the recovery of component luminosities as a function of redshift, if Ms observations are already present (\S\ref{sec:multi}).}
\label{fig:redshiftmulti_TestNs}
\end{center}
\end{figure}

Finally, we examine the efficacy of Ns in combinations of instruments observing over $0<z<5$ (Figure \ref{fig:redshiftmulti_TestNs}). At most redshifts, the inclusion of Ns observations do not significantly enhance the recovery of total or component luminosities. 

\section{Discussion and Conclusions}\label{sec:discCon}
We have studied the recovery of rest-frame $1-1000\mu$m  total, starburst, AGN, and host luminosities of infrared-luminous galaxies via SED fitting of photometric and low-resolution spectroscopic data. We have considered six fiducial near- to far-infrared instruments with real-world analogues, but have kept the simulations agnostic to specific facilities or survey designs.

\subsection{Luminosity recovery}
The recovery of total and component luminosities of infrared-luminous galaxies depends on several interlinked factors.  For $L_{Tot}$, bias is minimized with observations over the broadest possible wavelength range. Spectroscopic, rather than photometric, coverage helps to minimize random error.  For determining starburst, AGN, and host luminosities,  the primary requirement is dense wavelength sampling where the component SED peaks - far-infrared for starburst, mid-infrared for AGN, and near-infrared for host. These wavelengths also correspond to where each component shows the greatest range in SED shapes. Thus, observations with different instruments will give different levels of accuracy and precision in recovered luminosities. Observations at far-infrared wavelengths will recover starburst luminosities that are either unbiased, or biased high, but recover AGN luminosities that are biased low. Observations that favour mid-infrared wavelengths will give the opposite - unbiased or slightly high biased AGN luminosities, and starburst luminosities that are biased low. However, while we find that starburst and AGN luminosities can be recovered with minimal bias using appropriate observations, the recovered host galaxy luminosities are almost always biased low, by a factor of about two, on average.  A conceptually similar result has been found in previous studies \citep{mitchell13,sorba15,li21rmsloan}. We ascribe this to the challenge of measuring a faint host SED against the much brighter starburst/AGN SEDs.  

The specific instruments that give the best luminosity recoveries depend on several factors.  For a randomly selected source at $z\lesssim0.1$, the best recovery of starburst luminosity is given by observations that include those at far-infrared wavelengths, especially those with dense wavelength sampling. For AGN luminosity, the best recovery is achieved with mid-infrared observations, with dense wavelength sampling again having a positive effect. Host galaxy luminosities are best recovered with observations that include the near-infrared, though with the persistent low bias noted above.  If the source is selected to be starburst (AGN) dominated, then the observations needed to achieve the least biased recovery of luminosities do not change, but the level of bias in the recovery does. Starburst (AGN) dominated systems (as distinguished on the basis of $6.2\mu$m PAH EW, see \S\ref{sec:paheffect}) tend to show more (less) systematic bias for all of starburst, AGN, and host luminosity. The typical change in systematic bias relative to a randomly selected object is about 20\% in both cases. 

The S/N of the observations plays a significant role. With S/N$\lesssim6$, the precision starts to degrade, with accuracy starting to degrade at S/N$\lesssim4$. With S/Ns of $6-10$, the accuracy and percision both improve, and with S/Ns of $10-20$ there is little change in either.  

Angular resolution also plays an important role. Compared to the unresolved case, observing at (sub-)arcsecond resolution reduces both systematic bias and random error for all mid- through far-infrared observations. Starburst luminosities can be recovered with almost no systematic bias with mid-infrared observations, while AGN luminosity recovery becomes feasible with far-infrared observations, albeit only those with dense wavelength sampling. Since we neglect source confusion, the true benefit of observing at finer angular resolution is likely greater than we present here. 

When considering the recovery of total and component luminosities as a function of redshift, we obtain broadly similar results, with mid-infrared observations most useful for AGN luminosity, and far-infrared observations most useful for starburst luminosity. However, these trends evolve with redshift. AGN luminosity recovery using observed-frame mid-infrared observations becomes progressively more biased low as redshift increases. Starburst luminosity recovery with far-infrared observations, however, shows much less increase in systematic bias, at least up to $z\sim5$. Furthermore, as redshift increases, far-infrared observations become progressively better at measuring AGN luminosity. At most redshifts, we again see that starburst dominated systems show more biased luminosity recovery than do AGN dominated systems. 

A potentially significant simplification in our analysis was the exclusion of mid- to far-infrared fine-structure lines that are commonly observed in LIRG spectra (primarily those from Neon, Sulfur, Oxygen, Nitrogen, and Carbon) from the input SEDs to the simulations.  Since these lines are not present in the SEDs we fit to the simulated data, their inclusion in the input SEDs would have some inmpact on our results.  We defer an analysis of this scenario to a future work, and here only briefly speculate on the magnitude of this impact. For the spectrographs (Ns, Ms, and Fs) the impact is likely insignificant as the regions of continuum that contain these lines can be identified and removed.  For the NMp and Fp instruments though, this is not possible, so the impact will be larger. There will be some systematic bias high (since the fits would ascribe the line and continuum flux to continuum) with additional random error caused by variation in line strengths for a given starburst or AGN luminosity.  We speculate though, based on typical line luminosity to total infrared luminosity ratios of our sample \citep{far07,far13}, that the magnitude of both effects is of order 10\% or less.

\subsection{Implications for future work}
Several implications follow from these results.  First, studies that aim to calibrate quantities such as an infrared colour, or the flux of a mid- or far-infrared line, as a measure of luminosity, will be systematically biased unless the luminosities used to perform the calibration adhere to the requirements described above.  This is especially relevant for estimating the infrared luminosities of the $z\gtrsim4$ systems found by JWST, for which using calibrated relations is currently the only realistic route. It is likely also important that the population used to calibrate a line relation is similar in nature to the object under study.  Second is the outlook for future near- through far-infrared observatories. Our results show the importance of dense wavelength sampling, as well as wavelength coverage, at any wavelength. They also showcase the potentially transformative power of a far-infrared instrument with either multiple photometric bands, or, to an even greater degree, one that observes continuously from $20-30\mu$m to $200-300\mu$m.  Such an instrument reduces systematic bias and random error on the recovery of total and starburst luminosity over mid-infrared spectroscopy by at least a factor of four, and substantially enhances the recovery of AGN and host luminosity in conjunction with other instruments.   Either a multi-band far-infrared photometer or spectrometer covering a broad wavelength range is substantially better at recovering unbiased starburst and AGN luminosities than ground-based sub-millimter observations, at all redshifts. Examples of such far-infrared facilities include the SPIRIT interferometer, and the FIRESS and PRIMAger instruments onboard PRIMA. 

\begin{acknowledgments}
IS acknowledges fundings from the European Research Council (ERC) DistantDust (Grant No.101117541) and the Atraccíon de Talento Grant No.2022-T1/TIC-20472 of the Comunidad de Madrid, Spain. MB acknowledges support from the INAF minigrant “A systematic search for ultra-bright high-z strongly lensed galaxies in Planck catalogues”. DLC acknowledges support from UK Space Agency/STFC grant ST/Y006097/1. EH gratefully acknowledges the hospitality of the IAC during her long research visit and while this paper was being written. DBS gratefully acknowledges support from NSF Grant 2407752.  The National Radio Astronomy Observatory and Green Bank Observatory are facilities of the U.S. National Science Foundation operated under cooperative agreement by Associated Universities, Inc. We thank River, John Petty, and Alaine Lee for useful discussions. 
\end{acknowledgments}

\bibliography{simrecover}{}
\bibliographystyle{aasjournalv7}

\appendix
\restartappendixnumbering

\section{True to Observed ratios}\label{app:numbers}
We here present tables of the true-to-observed luminosity ratios derived and their $1\sigma$ uncertainties, for single- (Table \ref{tbl:highzsingle}) and multi-instrument observations (Table \ref{tbl:localmulti}).

\startlongtable
\begin{deluxetable}{rlccccccc}
\tablecaption{The observed-to-true ratios, $\alpha$ (Equation \ref{eq:alphadef}) for total, starburst, AGN, and host luminosity using observations with single instruments. Results are presented for $z\sim0$, $z=0.5$, $z=1$, $z=2$, $z=3$, $z=4$, and $z=5$. For each redshift, results are presented for simulations using the entire input sample, and for high and low PAH $6.2\mu$m EW sub-samples. For each of these samples, results are presented assuming the observations are spatially unresolved, and at angular resolutions of $1\arcsec$ for starburst and AGN, and $0.1\arcsec$ for AGN.}
\label{tbl:highzsingle}
\tablehead{
\colhead{} & 
\colhead{} &  
\multicolumn{4}{c}{Unresolved} & 
\multicolumn{2}{c}{$1.0\arcsec$}  & 
\colhead{$0.1\arcsec$}  \\
\colhead{Instrument} & 
\colhead{$W_{6.2}$} &  
\colhead{$\alpha_{Tot}$} & 
\colhead{$\alpha_{Sb}$} & 
\colhead{$\alpha_{AGN}$} & 
\colhead{$\alpha_{H}$} &
\colhead{$\alpha_{Sb}$} & 
\colhead{$\alpha_{AGN}$} & 
\colhead{$\alpha_{AGN}$} 
}
\startdata
\multicolumn{9}{l}{$z\sim0$} \\
\hline   
Ns  & All  & $0.30^{+0.66}_{-0.20}$  & $0.22^{+0.86}_{-0.19}$  & $0.32^{+0.90}_{-0.24}$  & $0.68^{+0.51}_{-0.28}$  & $0.74^{+0.84}_{-0.43}$  & $0.06^{+1.14}_{-0.06}$  & $0.41^{+0.88}_{-0.40}$ \\ 
  & $>0.14$  & $0.21^{+0.35}_{-0.13}$  & $0.14^{+0.38}_{-0.12}$  & $0.22^{+0.77}_{-0.15}$  & $0.82^{+0.76}_{-0.37}$  & $0.61^{+0.48}_{-0.28}$  & $0.01^{+0.38}_{-0.01}$  & $0.24^{+0.66}_{-0.20}$ \\ 
  & $<0.14$  & $0.42^{+1.19}_{-0.28}$  & $0.35^{+2.30}_{-0.31}$  & $0.42^{+0.87}_{-0.35}$  & $0.63^{+0.29}_{-0.30}$  & $1.21^{+0.57}_{-0.95}$  & $0.38^{+1.03}_{-0.35}$  & $0.75^{+0.75}_{-0.74}$ \\ 
Ms  & All  & $0.55^{+0.46}_{-0.21}$  & $0.29^{+0.80}_{-0.25}$  & $1.18^{+0.61}_{-0.21}$  & $0.42^{+0.53}_{-0.16}$  & $0.95^{+0.30}_{-0.31}$  & $1.26^{+0.29}_{-0.25}$  & $1.04^{+0.13}_{-0.49}$ \\ 
  & $>0.14$  & $0.40^{+0.13}_{-0.12}$  & $0.24^{+0.22}_{-0.20}$  & $1.29^{+0.50}_{-0.30}$  & $0.37^{+0.67}_{-0.11}$  & $0.75^{+0.28}_{-0.13}$  & $1.31^{+0.23}_{-0.20}$  & $1.04^{+0.11}_{-0.29}$ \\ 
  & $<0.14$  & $0.82^{+0.25}_{-0.29}$  & $0.49^{+0.89}_{-0.45}$  & $1.15^{+0.55}_{-0.38}$  & $0.47^{+0.44}_{-0.23}$  & $1.08^{+0.20}_{-0.27}$  & $1.15^{+0.43}_{-0.45}$  & $1.04^{+0.13}_{-0.92}$ \\ 
NMp  & All  & $0.45^{+0.58}_{-0.23}$  & $0.20^{+0.80}_{-0.17}$  & $1.23^{+0.91}_{-0.49}$  & $0.49^{+0.56}_{-0.20}$  & $0.44^{+0.94}_{-0.20}$  & $1.19^{+0.43}_{-0.55}$  & $1.07^{+0.38}_{-0.35}$ \\ 
  & $>0.14$  & $0.28^{+0.20}_{-0.09}$  & $0.10^{+0.25}_{-0.08}$  & $1.32^{+1.02}_{-0.54}$  & $0.45^{+0.65}_{-0.18}$  & $0.32^{+0.34}_{-0.10}$  & $1.28^{+0.49}_{-0.49}$  & $1.09^{+0.39}_{-0.22}$ \\ 
  & $<0.14$  & $0.73^{+0.60}_{-0.35}$  & $0.42^{+1.31}_{-0.35}$  & $1.16^{+0.83}_{-0.48}$  & $0.52^{+0.49}_{-0.17}$  & $0.83^{+0.91}_{-0.52}$  & $1.15^{+0.34}_{-0.85}$  & $1.06^{+0.35}_{-0.83}$ \\ 
Fs  & All  & $1.07^{+0.19}_{-0.10}$  & $1.24^{+0.36}_{-0.12}$  & $0.53^{+1.17}_{-0.34}$  & $0.51^{+0.53}_{-0.25}$  & $1.05^{+0.06}_{-0.08}$  & $0.72^{+0.73}_{-0.62}$  & $1.23^{+0.51}_{-0.37}$ \\ 
  & $>0.14$  & $1.05^{+0.07}_{-0.07}$  & $1.21^{+0.18}_{-0.08}$  & $0.29^{+0.38}_{-0.16}$  & $0.45^{+0.66}_{-0.21}$  & $1.04^{+0.05}_{-0.05}$  & $0.35^{+0.83}_{-0.25}$  & $1.35^{+0.61}_{-0.42}$ \\ 
  & $<0.14$  & $1.11^{+0.39}_{-0.15}$  & $1.30^{+0.40}_{-0.19}$  & $1.03^{+1.09}_{-0.65}$  & $0.58^{+0.41}_{-0.31}$  & $1.07^{+0.05}_{-0.13}$  & $1.08^{+0.52}_{-0.94}$  & $1.17^{+0.39}_{-0.39}$ \\ 
Fp  & All  & $1.50^{+0.60}_{-0.40}$  & $1.52^{+0.44}_{-0.32}$  & $2.08^{+4.57}_{-1.65}$  & $0.51^{+0.56}_{-0.25}$  & $1.40^{+0.26}_{-0.33}$  & $1.11^{+1.98}_{-0.93}$  & $0.97^{+0.74}_{-0.61}$ \\ 
  & $>0.14$  & $1.59^{+0.51}_{-0.30}$  & $1.60^{+0.35}_{-0.25}$  & $2.52^{+4.96}_{-2.01}$  & $0.46^{+0.61}_{-0.20}$  & $1.48^{+0.22}_{-0.17}$  & $1.45^{+2.44}_{-1.00}$  & $0.96^{+0.71}_{-0.54}$ \\ 
  & $<0.14$  & $1.36^{+0.74}_{-0.39}$  & $1.45^{+0.56}_{-0.33}$  & $1.69^{+4.30}_{-1.31}$  & $0.56^{+0.53}_{-0.28}$  & $1.24^{+0.34}_{-0.28}$  & $0.85^{+1.54}_{-0.82}$  & $0.97^{+0.80}_{-0.72}$ \\ 
\hline
\multicolumn{9}{l}{z=0.5} \\
\hline
Ms  & All  & $0.48^{+0.52}_{-0.21}$  & $0.24^{+0.78}_{-0.22}$  & $1.17^{+0.43}_{-0.52}$  & $0.35^{+0.56}_{-0.18}$  & $0.79^{+0.30}_{-0.35}$  & $1.19^{+0.56}_{-0.40}$  & $1.14^{+0.29}_{-0.29}$ \\ 
  & $>0.14$  & $0.37^{+0.14}_{-0.15}$  & $0.25^{+0.20}_{-0.18}$  & $1.16^{+0.47}_{-0.45}$  & $0.32^{+0.67}_{-0.15}$  & $0.58^{+0.24}_{-0.29}$  & $1.23^{+0.52}_{-0.39}$  & $1.17^{+0.36}_{-0.29}$ \\ 
  & $<0.14$  & $0.75^{+0.34}_{-0.31}$  & $0.24^{+1.08}_{-0.23}$  & $1.18^{+0.40}_{-0.58}$  & $0.38^{+0.45}_{-0.20}$  & $0.97^{+0.20}_{-0.24}$  & $1.19^{+0.56}_{-0.69}$  & $1.12^{+0.23}_{-0.65}$ \\ 
NMp  & All  & $0.30^{+0.61}_{-0.15}$  & $0.05^{+0.57}_{-0.04}$  & $1.08^{+0.60}_{-0.60}$  & $0.70^{+0.58}_{-0.44}$  & $0.32^{+0.68}_{-0.17}$  & $1.05^{+0.69}_{-0.65}$  & $0.96^{+0.39}_{-0.49}$ \\ 
  & $>0.14$  & $0.19^{+0.10}_{-0.09}$  & $0.02^{+0.05}_{-0.02}$  & $1.10^{+0.68}_{-0.53}$  & $0.58^{+0.86}_{-0.34}$  & $0.20^{+0.49}_{-0.08}$  & $0.97^{+0.71}_{-0.49}$  & $0.97^{+0.46}_{-0.45}$ \\ 
  & $<0.14$  & $0.63^{+0.57}_{-0.37}$  & $0.28^{+1.10}_{-0.28}$  & $1.08^{+0.49}_{-0.71}$  & $0.78^{+0.45}_{-0.52}$  & $0.62^{+0.77}_{-0.40}$  & $1.11^{+0.74}_{-0.77}$  & $0.94^{+0.37}_{-0.65}$ \\ 
Fs  & All  & $1.02^{+0.09}_{-0.14}$  & $1.20^{+0.28}_{-0.26}$  & $0.58^{+0.51}_{-0.51}$  & $0.46^{+0.72}_{-0.29}$  & $0.93^{+0.14}_{-0.66}$  & $0.57^{+0.88}_{-0.51}$  & $1.02^{+0.23}_{-0.42}$ \\ 
  & $>0.14$  & $1.02^{+0.07}_{-0.27}$  & $1.18^{+0.21}_{-0.26}$  & $0.38^{+0.49}_{-0.34}$  & $0.31^{+0.70}_{-0.15}$  & $0.62^{+0.40}_{-0.47}$  & $0.18^{+0.85}_{-0.14}$  & $0.94^{+0.27}_{-0.37}$ \\ 
  & $<0.14$  & $1.02^{+0.12}_{-0.13}$  & $1.22^{+0.35}_{-0.23}$  & $0.85^{+0.40}_{-0.69}$  & $0.56^{+0.67}_{-0.34}$  & $1.01^{+0.07}_{-0.48}$  & $0.86^{+0.70}_{-0.75}$  & $1.05^{+0.21}_{-0.38}$ \\ 
Fp  & All  & $1.33^{+0.38}_{-0.34}$  & $1.46^{+0.37}_{-0.27}$  & $1.36^{+2.70}_{-1.11}$  & $0.46^{+0.74}_{-0.24}$  & $1.27^{+0.27}_{-0.24}$  & $1.84^{+3.09}_{-1.51}$  & $0.94^{+1.10}_{-0.71}$ \\ 
  & $>0.14$  & $1.43^{+0.28}_{-0.21}$  & $1.51^{+0.32}_{-0.20}$  & $1.84^{+2.96}_{-1.26}$  & $0.36^{+0.81}_{-0.15}$  & $1.36^{+0.23}_{-0.18}$  & $2.30^{+3.25}_{-1.47}$  & $0.74^{+1.12}_{-0.48}$ \\ 
  & $<0.14$  & $1.15^{+0.55}_{-0.26}$  & $1.40^{+0.43}_{-0.33}$  & $0.97^{+2.01}_{-0.80}$  & $0.56^{+0.67}_{-0.32}$  & $1.17^{+0.30}_{-0.26}$  & $1.30^{+2.93}_{-1.18}$  & $1.14^{+1.04}_{-0.97}$ \\   
\hline
\multicolumn{9}{l}{z=1} \\
\hline
Ms  & All  & $0.55^{+0.50}_{-0.33}$  & $0.19^{+1.01}_{-0.18}$  & $1.31^{+0.81}_{-0.90}$  & $0.18^{+0.32}_{-0.10}$  & $0.73^{+0.78}_{-0.37}$  & $1.24^{+1.48}_{-0.85}$  & $1.16^{+1.02}_{-0.44}$ \\ 
  & $>0.14$  & $0.26^{+0.18}_{-0.09}$  & $0.03^{+0.24}_{-0.03}$  & $1.33^{+0.63}_{-0.79}$  & $0.15^{+0.17}_{-0.07}$  & $0.46^{+0.48}_{-0.13}$  & $1.17^{+1.33}_{-0.71}$  & $1.15^{+1.01}_{-0.39}$ \\ 
  & $<0.14$  & $0.92^{+0.39}_{-0.33}$  & $0.30^{+1.45}_{-0.29}$  & $1.29^{+0.88}_{-0.93}$  & $0.19^{+0.58}_{-0.17}$  & $1.02^{+0.72}_{-0.46}$  & $1.41^{+1.31}_{-1.37}$  & $1.19^{+1.10}_{-0.88}$ \\ 
NMp  & All  & $0.27^{+0.45}_{-0.13}$  & $0.07^{+0.36}_{-0.07}$  & $0.88^{+0.59}_{-0.66}$  & $0.72^{+0.72}_{-0.39}$  & $0.30^{+0.53}_{-0.21}$  & $0.75^{+0.72}_{-0.46}$  & $0.81^{+0.49}_{-0.41}$ \\ 
  & $>0.14$  & $0.16^{+0.14}_{-0.05}$  & $0.03^{+0.14}_{-0.03}$  & $0.73^{+0.81}_{-0.42}$  & $0.71^{+0.87}_{-0.41}$  & $0.22^{+0.15}_{-0.13}$  & $0.78^{+0.81}_{-0.45}$  & $0.79^{+0.54}_{-0.29}$ \\ 
  & $<0.14$  & $0.50^{+0.51}_{-0.26}$  & $0.17^{+0.81}_{-0.17}$  & $0.93^{+0.49}_{-0.75}$  & $0.72^{+0.51}_{-0.32}$  & $0.50^{+0.55}_{-0.41}$  & $0.72^{+0.69}_{-0.49}$  & $0.85^{+0.44}_{-0.71}$ \\ 
Fs  & All  & $1.01^{+0.07}_{-0.16}$  & $1.12^{+0.26}_{-0.27}$  & $0.74^{+0.44}_{-0.73}$  & $0.28^{+0.68}_{-0.21}$  & $0.94^{+0.14}_{-0.85}$  & $0.66^{+0.65}_{-0.62}$  & $0.86^{+0.28}_{-0.36}$ \\ 
  & $>0.14$  & $1.01^{+0.09}_{-0.21}$  & $1.12^{+0.18}_{-0.33}$  & $0.28^{+0.69}_{-0.27}$  & $0.18^{+0.33}_{-0.12}$  & $0.78^{+0.27}_{-0.69}$  & $0.21^{+0.71}_{-0.21}$  & $0.82^{+0.17}_{-0.28}$ \\ 
  & $<0.14$  & $1.01^{+0.07}_{-0.14}$  & $1.12^{+0.33}_{-0.19}$  & $0.84^{+0.42}_{-0.62}$  & $0.60^{+0.62}_{-0.49}$  & $1.00^{+0.13}_{-0.52}$  & $0.94^{+0.57}_{-0.80}$  & $0.99^{+0.17}_{-0.58}$ \\ 
Fp  & All  & $1.21^{+0.28}_{-0.22}$  & $1.36^{+0.45}_{-0.25}$  & $0.96^{+1.31}_{-0.74}$  & $0.42^{+0.82}_{-0.27}$  & $1.14^{+0.19}_{-0.18}$  & $1.14^{+1.75}_{-0.87}$  & $0.98^{+1.15}_{-0.66}$ \\ 
  & $>0.14$  & $1.25^{+0.21}_{-0.16}$  & $1.41^{+0.40}_{-0.24}$  & $0.90^{+1.21}_{-0.68}$  & $0.31^{+0.48}_{-0.16}$  & $1.19^{+0.18}_{-0.14}$  & $1.13^{+1.83}_{-0.78}$  & $0.80^{+1.06}_{-0.50}$ \\ 
  & $<0.14$  & $1.16^{+0.41}_{-0.29}$  & $1.30^{+0.52}_{-0.27}$  & $1.03^{+1.34}_{-0.80}$  & $0.61^{+0.82}_{-0.43}$  & $1.10^{+0.19}_{-0.19}$  & $1.15^{+1.66}_{-0.99}$  & $1.12^{+1.16}_{-0.80}$ \\   
\hline
\multicolumn{9}{l}{z=2} \\
\hline
Ms  & All  & $0.61^{+0.65}_{-0.36}$  & $0.34^{+1.21}_{-0.34}$  & $0.88^{+0.55}_{-0.85}$  & $0.27^{+0.75}_{-0.26}$  & $0.73^{+0.39}_{-0.44}$  & $1.20^{+0.44}_{-0.45}$  & $1.14^{+0.41}_{-0.28}$ \\ 
  & $>0.14$  & $0.35^{+0.50}_{-0.17}$  & $0.34^{+0.54}_{-0.34}$  & $0.55^{+0.65}_{-0.54}$  & $0.42^{+0.81}_{-0.41}$  & $0.41^{+0.13}_{-0.14}$  & $1.18^{+0.59}_{-0.41}$  & $1.14^{+0.43}_{-0.24}$ \\ 
  & $<0.14$  & $0.85^{+0.73}_{-0.41}$  & $0.41^{+1.83}_{-0.41}$  & $1.06^{+0.44}_{-0.79}$  & $0.20^{+0.67}_{-0.19}$  & $1.01^{+0.19}_{-0.29}$  & $1.20^{+0.37}_{-0.82}$  & $1.14^{+0.29}_{-0.66}$ \\ 
NMp  & All  & $0.63^{+0.90}_{-0.40}$  & $0.60^{+1.87}_{-0.55}$  & $0.74^{+0.87}_{-0.58}$  & $0.74^{+0.69}_{-0.43}$  & $0.03^{+0.28}_{-0.03}$  & $1.01^{+0.71}_{-0.57}$  & $0.97^{+0.58}_{-0.51}$ \\ 
  & $>0.14$  & $0.40^{+0.36}_{-0.27}$  & $0.39^{+0.48}_{-0.37}$  & $0.83^{+0.93}_{-0.64}$  & $0.65^{+0.99}_{-0.35}$  & $0.01^{+0.05}_{-0.01}$  & $1.01^{+0.88}_{-0.49}$  & $1.03^{+0.57}_{-0.36}$ \\ 
  & $<0.14$  & $0.91^{+1.40}_{-0.49}$  & $1.14^{+2.88}_{-1.05}$  & $0.63^{+0.86}_{-0.50}$  & $0.82^{+0.49}_{-0.49}$  & $0.11^{+0.48}_{-0.11}$  & $1.02^{+0.61}_{-0.72}$  & $0.92^{+0.56}_{-0.58}$ \\ 
Fs  & All  & $1.03^{+0.08}_{-0.11}$  & $1.12^{+0.29}_{-0.21}$  & $1.05^{+0.67}_{-0.67}$  & $0.27^{+0.99}_{-0.25}$  & $0.89^{+0.18}_{-0.53}$  & $0.44^{+0.94}_{-0.39}$  & $0.95^{+0.32}_{-0.55}$ \\ 
  & $>0.14$  & $0.99^{+0.10}_{-0.08}$  & $1.05^{+0.21}_{-0.13}$  & $1.07^{+2.01}_{-0.60}$  & $0.25^{+0.87}_{-0.21}$  & $0.88^{+0.13}_{-0.44}$  & $0.22^{+0.51}_{-0.20}$  & $0.84^{+0.26}_{-0.45}$ \\ 
  & $<0.14$  & $1.05^{+0.08}_{-0.08}$  & $1.19^{+0.45}_{-0.32}$  & $1.00^{+0.52}_{-0.66}$  & $0.29^{+1.10}_{-0.28}$  & $0.93^{+0.17}_{-0.57}$  & $0.90^{+0.85}_{-0.80}$  & $1.06^{+0.32}_{-0.63}$ \\ 
Fp  & All  & $1.18^{+0.17}_{-0.14}$  & $1.40^{+0.38}_{-0.23}$  & $0.81^{+1.01}_{-0.63}$  & $0.35^{+0.79}_{-0.23}$  & $1.27^{+0.30}_{-0.23}$  & $1.72^{+3.21}_{-1.41}$  & $1.01^{+1.16}_{-0.83}$ \\ 
  & $>0.14$  & $1.17^{+0.14}_{-0.13}$  & $1.38^{+0.25}_{-0.19}$  & $0.59^{+0.72}_{-0.43}$  & $0.37^{+0.64}_{-0.21}$  & $1.38^{+0.32}_{-0.21}$  & $2.22^{+4.49}_{-1.73}$  & $0.72^{+1.02}_{-0.54}$ \\ 
  & $<0.14$  & $1.18^{+0.19}_{-0.15}$  & $1.43^{+0.57}_{-0.30}$  & $1.15^{+0.92}_{-0.94}$  & $0.35^{+1.07}_{-0.25}$  & $1.17^{+0.23}_{-0.24}$  & $1.52^{+2.12}_{-1.34}$  & $1.30^{+1.18}_{-1.14}$ \\ 
\hline
\multicolumn{9}{l}{z=3} \\
\hline
Ms  & All  & $0.58^{+0.93}_{-0.37}$  & $0.64^{+1.37}_{-0.64}$  & $0.36^{+0.95}_{-0.35}$  & $0.85^{+0.71}_{-0.76}$  & $0.72^{+0.70}_{-0.70}$  & $1.20^{+0.89}_{-0.61}$  & $1.14^{+0.68}_{-0.54}$ \\ 
  & $>0.14$  & $0.33^{+0.42}_{-0.15}$  & $0.23^{+0.74}_{-0.23}$  & $0.45^{+0.91}_{-0.41}$  & $0.85^{+0.66}_{-0.43}$  & $0.32^{+0.45}_{-0.32}$  & $1.37^{+0.76}_{-0.76}$  & $1.41^{+0.69}_{-0.46}$ \\ 
  & $<0.14$  & $1.06^{+1.42}_{-0.62}$  & $1.58^{+1.78}_{-1.45}$  & $0.23^{+1.05}_{-0.22}$  & $0.78^{+0.93}_{-0.71}$  & $1.16^{+0.54}_{-0.53}$  & $1.16^{+0.28}_{-0.75}$  & $1.03^{+0.36}_{-0.44}$ \\ 
NMp  & All  & $0.76^{+1.63}_{-0.54}$  & $0.70^{+2.61}_{-0.68}$  & $0.42^{+1.26}_{-0.39}$  & $0.60^{+1.04}_{-0.46}$  & $0.07^{+0.27}_{-0.07}$  & $0.83^{+0.65}_{-0.48}$  & $0.91^{+0.42}_{-0.50}$ \\ 
  & $>0.14$  & $0.50^{+0.84}_{-0.39}$  & $0.45^{+1.02}_{-0.45}$  & $0.32^{+1.36}_{-0.32}$  & $0.54^{+0.83}_{-0.48}$  & $0.09^{+0.19}_{-0.08}$  & $0.76^{+0.66}_{-0.38}$  & $0.86^{+0.52}_{-0.32}$ \\ 
  & $<0.14$  & $1.11^{+1.66}_{-0.75}$  & $1.45^{+2.63}_{-1.39}$  & $0.60^{+1.09}_{-0.54}$  & $0.67^{+1.07}_{-0.47}$  & $0.04^{+0.41}_{-0.04}$  & $0.91^{+0.58}_{-0.71}$  & $0.96^{+0.34}_{-0.90}$ \\ 
Fs  & All  & $1.00^{+0.11}_{-0.11}$  & $1.07^{+0.23}_{-0.20}$  & $0.99^{+0.37}_{-0.43}$  & $0.17^{+0.51}_{-0.14}$  & $0.88^{+0.17}_{-0.71}$  & $0.76^{+0.88}_{-0.72}$  & $0.92^{+0.23}_{-0.37}$ \\ 
  & $>0.14$  & $0.94^{+0.08}_{-0.08}$  & $1.04^{+0.16}_{-0.13}$  & $0.92^{+0.27}_{-0.24}$  & $0.17^{+0.28}_{-0.12}$  & $0.80^{+0.21}_{-0.71}$  & $0.62^{+0.86}_{-0.59}$  & $0.88^{+0.21}_{-0.39}$ \\ 
  & $<0.14$  & $1.07^{+0.08}_{-0.12}$  & $1.13^{+0.36}_{-0.45}$  & $1.06^{+0.52}_{-0.74}$  & $0.13^{+1.18}_{-0.12}$  & $0.93^{+0.14}_{-0.48}$  & $0.80^{+0.95}_{-0.75}$  & $0.99^{+0.18}_{-0.36}$ \\ 
Fp  & All  & $1.12^{+0.15}_{-0.14}$  & $1.32^{+0.39}_{-0.28}$  & $0.77^{+0.95}_{-0.55}$  & $0.23^{+0.96}_{-0.16}$  & $1.21^{+0.21}_{-0.20}$  & $1.11^{+2.02}_{-0.93}$  & $1.17^{+1.10}_{-0.94}$ \\ 
  & $>0.14$  & $1.08^{+0.13}_{-0.12}$  & $1.29^{+0.22}_{-0.21}$  & $0.61^{+0.62}_{-0.38}$  & $0.19^{+0.35}_{-0.12}$  & $1.20^{+0.16}_{-0.17}$  & $1.31^{+2.03}_{-1.02}$  & $0.98^{+1.40}_{-0.68}$ \\ 
  & $<0.14$  & $1.16^{+0.18}_{-0.15}$  & $1.37^{+0.66}_{-0.43}$  & $1.07^{+1.01}_{-0.84}$  & $0.27^{+2.18}_{-0.21}$  & $1.22^{+0.27}_{-0.26}$  & $0.97^{+1.84}_{-0.90}$  & $1.24^{+1.00}_{-1.22}$ \\   
\hline
\multicolumn{9}{l}{z=4} \\
\hline
Ms  & All  & $0.63^{+0.90}_{-0.49}$  & $0.53^{+1.42}_{-0.53}$  & $0.17^{+0.99}_{-0.16}$  & $0.70^{+1.21}_{-0.63}$  & $0.77^{+0.59}_{-0.53}$  & $1.28^{+1.10}_{-0.82}$  & $1.07^{+0.61}_{-0.52}$ \\ 
  & $>0.14$  & $0.21^{+0.78}_{-0.12}$  & $0.19^{+0.89}_{-0.18}$  & $0.08^{+0.58}_{-0.08}$  & $0.70^{+0.58}_{-0.54}$  & $0.46^{+0.31}_{-0.30}$  & $1.62^{+0.79}_{-1.21}$  & $1.03^{+0.84}_{-0.37}$ \\ 
  & $<0.14$  & $0.85^{+2.33}_{-0.56}$  & $1.22^{+3.33}_{-1.22}$  & $0.49^{+1.01}_{-0.47}$  & $0.86^{+1.09}_{-0.83}$  & $1.09^{+0.45}_{-0.50}$  & $1.13^{+1.17}_{-0.66}$  & $1.15^{+0.49}_{-0.92}$ \\ 
NMp  & All  & $0.71^{+2.38}_{-0.59}$  & $0.57^{+4.43}_{-0.57}$  & $0.25^{+1.30}_{-0.23}$  & $0.39^{+1.28}_{-0.27}$  & $0.05^{+0.24}_{-0.05}$  & $0.81^{+0.83}_{-0.56}$  & $0.95^{+0.46}_{-0.52}$ \\ 
  & $>0.14$  & $0.45^{+1.44}_{-0.36}$  & $0.45^{+1.76}_{-0.45}$  & $0.26^{+0.86}_{-0.26}$  & $0.29^{+0.99}_{-0.19}$  & $0.03^{+0.11}_{-0.03}$  & $0.64^{+0.93}_{-0.38}$  & $0.86^{+0.42}_{-0.38}$ \\ 
  & $<0.14$  & $1.14^{+3.38}_{-1.01}$  & $0.91^{+6.62}_{-0.91}$  & $0.23^{+1.36}_{-0.19}$  & $0.59^{+1.38}_{-0.38}$  & $0.11^{+0.58}_{-0.11}$  & $0.97^{+0.70}_{-0.75}$  & $1.04^{+0.45}_{-0.78}$ \\ 
Fs  & All  & $0.98^{+0.10}_{-0.11}$  & $1.00^{+0.25}_{-0.18}$  & $1.11^{+0.30}_{-0.37}$  & $0.23^{+0.54}_{-0.19}$  & $0.74^{+0.27}_{-0.36}$  & $0.70^{+0.82}_{-0.68}$  & $0.98^{+0.24}_{-0.38}$ \\ 
  & $>0.14$  & $0.91^{+0.10}_{-0.07}$  & $0.95^{+0.18}_{-0.13}$  & $1.05^{+0.26}_{-0.29}$  & $0.27^{+0.41}_{-0.20}$  & $0.62^{+0.28}_{-0.54}$  & $0.32^{+0.91}_{-0.31}$  & $1.01^{+0.33}_{-0.44}$ \\ 
  & $<0.14$  & $1.03^{+0.07}_{-0.09}$  & $1.07^{+0.24}_{-0.25}$  & $1.16^{+0.32}_{-0.72}$  & $0.21^{+0.66}_{-0.19}$  & $0.92^{+0.16}_{-0.40}$  & $0.93^{+0.77}_{-0.82}$  & $0.94^{+0.24}_{-0.31}$ \\ 
Fp  & All  & $1.05^{+0.13}_{-0.12}$  & $1.24^{+0.41}_{-0.25}$  & $0.80^{+0.76}_{-0.61}$  & $0.21^{+0.78}_{-0.15}$  & $1.15^{+0.21}_{-0.19}$  & $1.07^{+1.95}_{-0.90}$  & $1.06^{+1.21}_{-0.83}$ \\ 
  & $>0.14$  & $1.01^{+0.10}_{-0.13}$  & $1.17^{+0.32}_{-0.18}$  & $0.55^{+0.71}_{-0.39}$  & $0.19^{+0.30}_{-0.13}$  & $1.20^{+0.22}_{-0.19}$  & $0.86^{+2.70}_{-0.63}$  & $0.92^{+1.10}_{-0.71}$ \\ 
  & $<0.14$  & $1.09^{+0.14}_{-0.12}$  & $1.32^{+0.53}_{-0.36}$  & $1.05^{+0.82}_{-0.79}$  & $0.25^{+1.30}_{-0.18}$  & $1.11^{+0.18}_{-0.19}$  & $1.26^{+1.43}_{-1.15}$  & $1.13^{+1.34}_{-0.89}$ \\ 
\hline
\multicolumn{9}{l}{z=5} \\
\hline
Ms  & All  & $0.43^{+1.22}_{-0.35}$  & $0.08^{+1.79}_{-0.08}$  & $0.10^{+1.28}_{-0.10}$  & $0.94^{+0.88}_{-0.67}$  & $0.78^{+0.35}_{-0.44}$  & $1.55^{+0.99}_{-0.86}$  & $1.20^{+0.64}_{-0.40}$ \\ 
  & $>0.14$  & $0.13^{+0.52}_{-0.05}$  & $0.02^{+0.72}_{-0.02}$  & $0.10^{+0.91}_{-0.10}$  & $1.00^{+2.26}_{-0.73}$  & $0.59^{+0.41}_{-0.30}$  & $1.56^{+0.81}_{-0.64}$  & $1.16^{+0.67}_{-0.30}$ \\ 
  & $<0.14$  & $1.00^{+1.93}_{-0.74}$  & $0.63^{+4.06}_{-0.63}$  & $0.16^{+3.58}_{-0.15}$  & $0.85^{+0.93}_{-0.59}$  & $0.89^{+0.37}_{-0.40}$  & $1.54^{+1.33}_{-1.08}$  & $1.26^{+1.01}_{-0.95}$ \\ 
NMp  & All  & $0.41^{+2.95}_{-0.34}$  & $0.03^{+3.01}_{-0.03}$  & $0.19^{+1.57}_{-0.19}$  & $0.51^{+1.02}_{-0.37}$  & $0.07^{+0.33}_{-0.06}$  & $0.78^{+0.56}_{-0.53}$  & $0.81^{+0.38}_{-0.45}$ \\ 
  & $>0.14$  & $0.40^{+1.83}_{-0.33}$  & $0.02^{+2.11}_{-0.02}$  & $0.19^{+2.76}_{-0.19}$  & $0.43^{+1.22}_{-0.28}$  & $0.03^{+0.15}_{-0.03}$  & $0.83^{+0.61}_{-0.49}$  & $0.75^{+0.32}_{-0.30}$ \\ 
  & $<0.14$  & $0.43^{+3.50}_{-0.35}$  & $0.09^{+4.96}_{-0.09}$  & $0.20^{+1.31}_{-0.18}$  & $0.55^{+0.83}_{-0.44}$  & $0.11^{+0.46}_{-0.10}$  & $0.76^{+0.44}_{-0.61}$  & $0.93^{+0.38}_{-0.86}$ \\ 
Fs  & All  & $0.96^{+0.14}_{-0.13}$  & $1.00^{+0.31}_{-0.30}$  & $1.09^{+0.35}_{-0.38}$  & $0.19^{+0.91}_{-0.13}$  & $0.79^{+0.25}_{-0.31}$  & $0.76^{+0.59}_{-0.68}$  & $0.99^{+0.23}_{-0.29}$ \\ 
  & $>0.14$  & $0.89^{+0.10}_{-0.14}$  & $0.93^{+0.17}_{-0.21}$  & $1.05^{+0.38}_{-0.30}$  & $0.14^{+0.26}_{-0.09}$  & $0.73^{+0.14}_{-0.45}$  & $0.53^{+0.62}_{-0.49}$  & $1.02^{+0.21}_{-0.30}$ \\ 
  & $<0.14$  & $1.05^{+0.13}_{-0.14}$  & $1.09^{+0.29}_{-0.52}$  & $1.13^{+0.34}_{-1.05}$  & $0.39^{+1.73}_{-0.31}$  & $0.91^{+0.19}_{-0.34}$  & $0.94^{+0.56}_{-0.85}$  & $0.98^{+0.23}_{-0.80}$ \\ 
Fp  & All  & $1.06^{+0.19}_{-0.14}$  & $1.24^{+0.44}_{-0.27}$  & $0.71^{+0.72}_{-0.56}$  & $0.21^{+0.81}_{-0.14}$  & $1.15^{+0.20}_{-0.18}$  & $1.32^{+2.48}_{-1.08}$  & $1.15^{+0.89}_{-0.81}$ \\ 
  & $>0.14$  & $1.03^{+0.18}_{-0.20}$  & $1.17^{+0.26}_{-0.30}$  & $0.67^{+0.54}_{-0.40}$  & $0.17^{+0.37}_{-0.11}$  & $1.16^{+0.16}_{-0.13}$  & $1.26^{+2.84}_{-1.00}$  & $1.10^{+1.06}_{-0.79}$ \\ 
  & $<0.14$  & $1.08^{+0.18}_{-0.11}$  & $1.37^{+0.71}_{-0.27}$  & $0.83^{+0.68}_{-0.78}$  & $0.44^{+0.80}_{-0.35}$  & $1.14^{+0.25}_{-0.24}$  & $1.33^{+2.27}_{-1.14}$  & $1.18^{+0.79}_{-0.76}$ \\ 
  \enddata
\end{deluxetable}

\startlongtable
\begin{deluxetable}{rlcccc}
\tablecaption{The observed-to-true ratios, $\alpha$ (Equation \ref{eq:alphadef}) for total, starburst, AGN, and host luminosity using observations with multiple instruments. Results are presented for $z\sim0$, $z=0.5$, $z=1$, $z=2$, $z=3$, $z=4$, and $z=5$. For each redshift, results are presented for simulations using the entire input sample, and for high and low PAH $6.2\mu$m EW sub-samples (\S\ref{sec:multi}). 
\label{tbl:localmulti}}
\tablehead{
\colhead{Instruments} & 
\colhead{$W_{6.2}$} &  
\colhead{$\alpha_{Tot}$} & 
\colhead{$\alpha_{Sb}$} & 
\colhead{$\alpha_{AGN}$} & 
\colhead{$\alpha_{H}$} 
}
\startdata
\multicolumn{6}{l}{$z\sim0$} \\
\hline
Ns Ms  & All  & $0.63^{+0.37}_{-0.26}$  & $0.37^{+0.58}_{-0.28}$  & $1.26^{+0.54}_{-0.32}$  & $0.93^{+0.60}_{-0.46}$ \\ 
  & $>0.14$  & $0.44^{+0.22}_{-0.14}$  & $0.28^{+0.22}_{-0.16}$  & $1.47^{+0.33}_{-0.48}$  & $0.98^{+0.54}_{-0.48}$ \\ 
  & $<0.14$  & $0.82^{+0.36}_{-0.24}$  & $0.55^{+0.75}_{-0.50}$  & $1.19^{+0.60}_{-0.30}$  & $0.86^{+0.71}_{-0.43}$ \\ 
Ns Fp  & All  & $1.11^{+0.22}_{-0.18}$  & $1.37^{+0.33}_{-0.23}$  & $0.23^{+0.96}_{-0.20}$  & $0.81^{+0.45}_{-0.40}$ \\ 
  & $>0.14$  & $1.14^{+0.17}_{-0.16}$  & $1.33^{+0.28}_{-0.18}$  & $0.13^{+0.32}_{-0.11}$  & $0.83^{+0.56}_{-0.41}$ \\ 
  & $<0.14$  & $1.08^{+0.28}_{-0.20}$  & $1.42^{+0.35}_{-0.31}$  & $0.50^{+0.92}_{-0.44}$  & $0.80^{+0.32}_{-0.41}$ \\ 
Ns Fs  & All  & $1.03^{+0.08}_{-0.09}$  & $1.19^{+0.24}_{-0.13}$  & $0.55^{+0.50}_{-0.47}$  & $0.79^{+0.35}_{-0.40}$ \\ 
  & $>0.14$  & $1.01^{+0.06}_{-0.08}$  & $1.17^{+0.18}_{-0.10}$  & $0.27^{+0.41}_{-0.21}$  & $0.73^{+0.49}_{-0.36}$ \\ 
  & $<0.14$  & $1.06^{+0.08}_{-0.11}$  & $1.23^{+0.25}_{-0.19}$  & $0.81^{+0.44}_{-0.67}$  & $0.81^{+0.29}_{-0.42}$ \\ 
Ns Sp  & All  & $1.42^{+0.94}_{-0.51}$  & $1.65^{+0.77}_{-1.07}$  & $0.21^{+3.52}_{-0.20}$  & $0.79^{+0.63}_{-0.35}$ \\ 
  & $>0.14$  & $1.62^{+1.04}_{-0.51}$  & $1.58^{+0.83}_{-0.76}$  & $0.23^{+6.34}_{-0.23}$  & $0.82^{+0.91}_{-0.33}$ \\ 
  & $<0.14$  & $1.31^{+0.65}_{-0.49}$  & $1.72^{+0.70}_{-1.34}$  & $0.18^{+1.60}_{-0.17}$  & $0.77^{+0.34}_{-0.38}$ \\ 
Ms Fp  & All  & $1.06^{+0.14}_{-0.13}$  & $1.05^{+0.21}_{-0.30}$  & $1.20^{+0.43}_{-0.33}$  & $0.48^{+0.48}_{-0.27}$ \\ 
  & $>0.14$  & $0.99^{+0.11}_{-0.12}$  & $1.00^{+0.18}_{-0.26}$  & $1.15^{+0.53}_{-0.35}$  & $0.59^{+0.48}_{-0.36}$ \\ 
  & $<0.14$  & $1.12^{+0.12}_{-0.12}$  & $1.11^{+0.26}_{-0.32}$  & $1.23^{+0.36}_{-0.33}$  & $0.45^{+0.41}_{-0.28}$ \\ 
Ms Fs  & All  & $1.04^{+0.09}_{-0.08}$  & $1.10^{+0.15}_{-0.17}$  & $1.11^{+0.27}_{-0.29}$  & $0.35^{+0.59}_{-0.18}$ \\ 
  & $>0.14$  & $1.01^{+0.07}_{-0.06}$  & $1.08^{+0.14}_{-0.15}$  & $1.00^{+0.39}_{-0.21}$  & $0.30^{+0.67}_{-0.14}$ \\ 
  & $<0.14$  & $1.08^{+0.07}_{-0.08}$  & $1.14^{+0.15}_{-0.18}$  & $1.12^{+0.25}_{-0.20}$  & $0.40^{+0.51}_{-0.14}$ \\ 
Ms Sp  & All  & $1.01^{+0.30}_{-0.60}$  & $0.96^{+0.56}_{-0.92}$  & $1.21^{+0.48}_{-0.32}$  & $0.51^{+0.47}_{-0.30}$ \\ 
  & $>0.14$  & $0.68^{+0.60}_{-0.41}$  & $0.63^{+0.68}_{-0.63}$  & $1.30^{+0.53}_{-0.30}$  & $0.46^{+0.48}_{-0.25}$ \\ 
  & $<0.14$  & $1.12^{+0.21}_{-0.24}$  & $1.22^{+0.52}_{-0.66}$  & $1.09^{+0.40}_{-0.24}$  & $0.58^{+0.42}_{-0.38}$ \\ 
NMp Fp  & All  & $1.13^{+0.13}_{-0.12}$  & $1.30^{+0.24}_{-0.19}$  & $0.91^{+0.46}_{-0.55}$  & $0.66^{+0.47}_{-0.40}$ \\ 
  & $>0.14$  & $1.14^{+0.11}_{-0.11}$  & $1.27^{+0.16}_{-0.15}$  & $0.83^{+0.52}_{-0.43}$  & $0.52^{+0.63}_{-0.29}$ \\ 
  & $<0.14$  & $1.13^{+0.17}_{-0.13}$  & $1.33^{+0.40}_{-0.24}$  & $0.97^{+0.39}_{-0.71}$  & $0.70^{+0.36}_{-0.39}$ \\ 
NMp Fs  & All  & $1.04^{+0.07}_{-0.07}$  & $1.20^{+0.20}_{-0.14}$  & $0.79^{+0.43}_{-0.52}$  & $0.47^{+0.52}_{-0.21}$ \\ 
  & $>0.14$  & $1.03^{+0.06}_{-0.07}$  & $1.18^{+0.15}_{-0.14}$  & $0.66^{+0.32}_{-0.44}$  & $0.39^{+0.65}_{-0.14}$ \\ 
  & $<0.14$  & $1.05^{+0.07}_{-0.08}$  & $1.21^{+0.35}_{-0.11}$  & $0.91^{+0.46}_{-0.32}$  & $0.60^{+0.36}_{-0.32}$ \\ 
NMp Sp  & All  & $1.59^{+0.76}_{-0.42}$  & $2.02^{+0.92}_{-0.77}$  & $0.89^{+0.79}_{-0.72}$  & $0.58^{+0.48}_{-0.28}$ \\ 
  & $>0.14$  & $1.88^{+0.76}_{-0.63}$  & $2.20^{+0.96}_{-0.80}$  & $0.79^{+0.98}_{-0.55}$  & $0.56^{+0.54}_{-0.25}$ \\ 
  & $<0.14$  & $1.43^{+0.44}_{-0.28}$  & $1.82^{+0.82}_{-0.66}$  & $0.97^{+0.67}_{-0.83}$  & $0.62^{+0.40}_{-0.34}$ \\ 
Ns Ms Fp  & All  & $1.05^{+0.15}_{-0.20}$  & $1.01^{+0.26}_{-0.38}$  & $1.22^{+0.53}_{-0.32}$  & $0.62^{+0.39}_{-0.26}$ \\ 
  & $>0.14$  & $0.97^{+0.19}_{-0.29}$  & $1.00^{+0.19}_{-0.79}$  & $1.40^{+0.42}_{-0.47}$  & $0.65^{+0.50}_{-0.33}$ \\ 
  & $<0.14$  & $1.10^{+0.13}_{-0.13}$  & $1.05^{+0.29}_{-0.27}$  & $1.16^{+0.53}_{-0.29}$  & $0.61^{+0.29}_{-0.23}$ \\ 
Ns Ms Fs  & All  & $1.04^{+0.08}_{-0.07}$  & $1.07^{+0.14}_{-0.11}$  & $1.07^{+0.23}_{-0.28}$  & $0.66^{+0.33}_{-0.31}$ \\ 
  & $>0.14$  & $1.02^{+0.05}_{-0.07}$  & $1.06^{+0.08}_{-0.07}$  & $1.03^{+0.20}_{-0.26}$  & $0.55^{+0.40}_{-0.22}$ \\ 
  & $<0.14$  & $1.10^{+0.04}_{-0.11}$  & $1.12^{+0.15}_{-0.17}$  & $1.09^{+0.24}_{-0.26}$  & $0.74^{+0.27}_{-0.26}$ \\ 
Ns Ms Sp  & All  & $0.85^{+0.35}_{-0.46}$  & $0.73^{+0.53}_{-0.72}$  & $1.23^{+0.47}_{-0.31}$  & $0.84^{+0.86}_{-0.45}$ \\ 
  & $>0.14$  & $0.49^{+0.43}_{-0.27}$  & $0.34^{+0.61}_{-0.34}$  & $1.35^{+0.48}_{-0.42}$  & $1.06^{+0.82}_{-0.73}$ \\ 
  & $<0.14$  & $1.03^{+0.20}_{-0.30}$  & $0.95^{+0.54}_{-0.40}$  & $1.15^{+0.32}_{-0.25}$  & $0.75^{+0.60}_{-0.34}$ \\ 
\hline
\multicolumn{6}{l}{z=0.5} \\
\hline 
Ms Fs  & All  & $0.97^{+0.11}_{-0.25}$  & $1.07^{+0.19}_{-0.40}$  & $0.87^{+0.42}_{-0.62}$  & $0.52^{+0.45}_{-0.37}$ \\ 
  & $>0.14$  & $0.92^{+0.10}_{-0.78}$  & $0.98^{+0.20}_{-0.98}$  & $0.69^{+0.41}_{-0.47}$  & $0.57^{+0.35}_{-0.43}$ \\ 
  & $<0.14$  & $1.02^{+0.09}_{-0.09}$  & $1.14^{+0.20}_{-0.28}$  & $0.94^{+0.52}_{-0.65}$  & $0.47^{+0.50}_{-0.31}$ \\ 
Ms Fp  & All  & $1.03^{+0.14}_{-0.13}$  & $1.06^{+0.24}_{-0.17}$  & $1.03^{+0.32}_{-0.35}$  & $0.58^{+0.50}_{-0.43}$ \\ 
  & $>0.14$  & $0.98^{+0.12}_{-0.10}$  & $1.03^{+0.21}_{-0.13}$  & $0.96^{+0.29}_{-0.22}$  & $0.73^{+0.42}_{-0.55}$ \\ 
  & $<0.14$  & $1.07^{+0.13}_{-0.16}$  & $1.10^{+0.25}_{-0.23}$  & $1.10^{+0.44}_{-0.64}$  & $0.42^{+0.63}_{-0.39}$ \\ 
Ms Sp  & All  & $1.00^{+0.20}_{-0.55}$  & $0.99^{+0.57}_{-0.77}$  & $1.02^{+0.35}_{-0.48}$  & $0.43^{+0.69}_{-0.30}$ \\ 
  & $>0.14$  & $0.63^{+0.48}_{-0.26}$  & $0.57^{+0.67}_{-0.35}$  & $1.02^{+0.43}_{-0.48}$  & $0.48^{+0.75}_{-0.35}$ \\ 
  & $<0.14$  & $1.03^{+0.18}_{-0.25}$  & $1.12^{+0.49}_{-0.82}$  & $1.02^{+0.28}_{-0.64}$  & $0.38^{+0.66}_{-0.23}$ \\ 
NMp Fs  & All  & $0.99^{+0.08}_{-0.17}$  & $1.11^{+0.32}_{-0.23}$  & $0.49^{+0.60}_{-0.42}$  & $0.66^{+0.62}_{-0.43}$ \\ 
  & $>0.14$  & $0.97^{+0.09}_{-0.15}$  & $1.08^{+0.19}_{-0.20}$  & $0.36^{+0.48}_{-0.35}$  & $0.68^{+0.39}_{-0.45}$ \\ 
  & $<0.14$  & $1.01^{+0.07}_{-0.16}$  & $1.15^{+0.47}_{-0.18}$  & $0.68^{+0.63}_{-0.49}$  & $0.66^{+0.68}_{-0.43}$ \\ 
NMp Fp  & All  & $1.09^{+0.15}_{-0.15}$  & $1.23^{+0.26}_{-0.21}$  & $0.82^{+0.54}_{-0.47}$  & $0.59^{+0.62}_{-0.37}$ \\ 
  & $>0.14$  & $1.13^{+0.17}_{-0.16}$  & $1.27^{+0.19}_{-0.24}$  & $0.81^{+0.58}_{-0.49}$  & $0.60^{+0.47}_{-0.39}$ \\ 
  & $<0.14$  & $1.06^{+0.10}_{-0.16}$  & $1.20^{+0.36}_{-0.19}$  & $0.83^{+0.53}_{-0.43}$  & $0.57^{+0.65}_{-0.34}$ \\ 
NMp Sp  & All  & $1.05^{+0.49}_{-0.49}$  & $1.15^{+0.93}_{-0.71}$  & $0.76^{+0.85}_{-0.54}$  & $0.74^{+0.65}_{-0.48}$ \\ 
  & $>0.14$  & $0.94^{+0.58}_{-0.70}$  & $1.04^{+0.65}_{-0.95}$  & $0.72^{+0.77}_{-0.45}$  & $0.80^{+0.69}_{-0.47}$ \\ 
  & $<0.14$  & $1.18^{+0.38}_{-0.34}$  & $1.39^{+0.86}_{-0.69}$  & $0.88^{+0.95}_{-0.74}$  & $0.68^{+0.63}_{-0.48}$ \\ 
  \hline
\multicolumn{6}{l}{z=1} \\
\hline 
Ms Fs  & All  & $0.95^{+0.14}_{-0.24}$  & $1.01^{+0.36}_{-0.76}$  & $0.83^{+0.65}_{-0.73}$  & $0.30^{+0.84}_{-0.30}$ \\ 
  & $>0.14$  & $0.87^{+0.15}_{-0.18}$  & $0.90^{+0.32}_{-0.79}$  & $0.74^{+0.73}_{-0.58}$  & $0.31^{+0.90}_{-0.24}$ \\ 
  & $<0.14$  & $1.00^{+0.12}_{-0.21}$  & $1.06^{+0.41}_{-0.58}$  & $0.88^{+0.58}_{-0.85}$  & $0.16^{+0.95}_{-0.16}$ \\ 
Ms Fp  & All  & $0.94^{+0.34}_{-0.28}$  & $0.78^{+0.59}_{-0.76}$  & $1.44^{+1.85}_{-0.59}$  & $0.17^{+0.58}_{-0.16}$ \\ 
  & $>0.14$  & $0.83^{+0.23}_{-0.36}$  & $0.74^{+0.42}_{-0.73}$  & $1.88^{+1.80}_{-0.50}$  & $0.24^{+0.53}_{-0.19}$ \\ 
  & $<0.14$  & $1.12^{+0.20}_{-0.33}$  & $1.15^{+0.34}_{-0.87}$  & $1.20^{+1.71}_{-0.72}$  & $0.11^{+0.60}_{-0.11}$ \\ 
Ms Sp  & All  & $0.94^{+0.33}_{-0.45}$  & $0.87^{+0.42}_{-0.83}$  & $1.10^{+1.56}_{-0.60}$  & $0.14^{+0.35}_{-0.14}$ \\ 
  & $>0.14$  & $0.65^{+0.58}_{-0.34}$  & $0.66^{+0.41}_{-0.49}$  & $1.09^{+1.54}_{-0.60}$  & $0.11^{+0.15}_{-0.11}$ \\ 
  & $<0.14$  & $1.04^{+0.41}_{-0.19}$  & $0.98^{+0.41}_{-0.97}$  & $1.11^{+1.56}_{-0.74}$  & $0.17^{+0.50}_{-0.17}$ \\ 
NMp Fs  & All  & $1.01^{+0.08}_{-0.10}$  & $1.17^{+0.26}_{-0.20}$  & $0.57^{+0.57}_{-0.53}$  & $0.77^{+0.83}_{-0.57}$ \\ 
  & $>0.14$  & $1.01^{+0.07}_{-0.10}$  & $1.16^{+0.13}_{-0.17}$  & $0.42^{+0.54}_{-0.38}$  & $0.47^{+1.05}_{-0.29}$ \\ 
  & $<0.14$  & $1.03^{+0.08}_{-0.11}$  & $1.18^{+0.43}_{-0.25}$  & $0.78^{+0.39}_{-0.75}$  & $0.96^{+0.65}_{-0.66}$ \\ 
NMp Fp  & All  & $1.03^{+0.17}_{-0.14}$  & $1.22^{+0.31}_{-0.19}$  & $0.65^{+0.43}_{-0.47}$  & $0.94^{+0.53}_{-0.59}$ \\ 
  & $>0.14$  & $1.06^{+0.18}_{-0.14}$  & $1.23^{+0.26}_{-0.20}$  & $0.49^{+0.50}_{-0.30}$  & $0.91^{+0.64}_{-0.69}$ \\ 
  & $<0.14$  & $1.01^{+0.14}_{-0.15}$  & $1.22^{+0.39}_{-0.18}$  & $0.74^{+0.46}_{-0.57}$  & $0.96^{+0.44}_{-0.48}$ \\ 
NMp Sp  & All  & $0.99^{+0.43}_{-0.35}$  & $1.22^{+0.74}_{-0.49}$  & $0.52^{+0.41}_{-0.39}$  & $0.91^{+0.93}_{-0.58}$ \\ 
  & $>0.14$  & $1.09^{+0.39}_{-0.43}$  & $1.23^{+0.70}_{-0.53}$  & $0.43^{+0.36}_{-0.26}$  & $0.77^{+1.34}_{-0.52}$ \\ 
  & $<0.14$  & $0.94^{+0.38}_{-0.31}$  & $1.20^{+0.78}_{-0.46}$  & $0.59^{+0.54}_{-0.48}$  & $0.96^{+0.82}_{-0.53}$ \\ 
\hline
\multicolumn{6}{l}{z=2} \\
\hline 
Ms Fs  & All  & $1.02^{+0.10}_{-0.09}$  & $0.98^{+0.27}_{-0.18}$  & $1.12^{+0.41}_{-0.33}$  & $1.15^{+1.04}_{-0.96}$ \\ 
  & $>0.14$  & $0.96^{+0.06}_{-0.06}$  & $0.97^{+0.10}_{-0.12}$  & $1.09^{+0.26}_{-0.26}$  & $1.38^{+0.69}_{-0.99}$ \\ 
  & $<0.14$  & $1.09^{+0.06}_{-0.13}$  & $1.04^{+0.25}_{-0.36}$  & $1.17^{+0.45}_{-0.89}$  & $1.01^{+1.30}_{-0.96}$ \\ 
Ms Fp  & All  & $0.96^{+0.19}_{-0.16}$  & $0.94^{+0.35}_{-0.39}$  & $1.09^{+0.45}_{-0.68}$  & $1.30^{+1.56}_{-1.02}$ \\ 
  & $>0.14$  & $0.92^{+0.08}_{-0.31}$  & $0.85^{+0.18}_{-0.23}$  & $1.14^{+0.62}_{-0.75}$  & $1.62^{+1.45}_{-0.94}$ \\ 
  & $<0.14$  & $1.04^{+0.20}_{-0.18}$  & $1.11^{+0.33}_{-0.57}$  & $1.07^{+0.31}_{-0.66}$  & $1.19^{+1.53}_{-1.04}$ \\ 
Ms Sp  & All  & $0.91^{+0.41}_{-0.29}$  & $0.96^{+0.53}_{-0.50}$  & $1.01^{+0.82}_{-0.46}$  & $0.87^{+1.25}_{-0.80}$ \\ 
  & $>0.14$  & $0.75^{+0.23}_{-0.27}$  & $0.67^{+0.34}_{-0.45}$  & $1.06^{+0.89}_{-0.48}$  & $1.19^{+1.33}_{-0.55}$ \\ 
  & $<0.14$  & $1.12^{+0.29}_{-0.31}$  & $1.18^{+0.52}_{-0.39}$  & $0.99^{+0.65}_{-0.51}$  & $0.51^{+1.21}_{-0.51}$ \\ 
NMp Fs  & All  & $1.03^{+0.08}_{-0.07}$  & $1.14^{+0.17}_{-0.24}$  & $0.92^{+0.51}_{-0.60}$  & $0.55^{+1.16}_{-0.34}$ \\ 
  & $>0.14$  & $1.01^{+0.09}_{-0.08}$  & $1.13^{+0.13}_{-0.21}$  & $0.58^{+0.86}_{-0.35}$  & $0.56^{+1.13}_{-0.39}$ \\ 
  & $<0.14$  & $1.04^{+0.07}_{-0.07}$  & $1.15^{+0.19}_{-0.27}$  & $1.02^{+0.38}_{-0.60}$  & $0.53^{+1.22}_{-0.30}$ \\ 
NMp Fp  & All  & $1.03^{+0.15}_{-0.15}$  & $1.25^{+0.33}_{-0.25}$  & $0.28^{+0.92}_{-0.19}$  & $0.90^{+0.90}_{-0.54}$ \\ 
  & $>0.14$  & $1.02^{+0.13}_{-0.13}$  & $1.17^{+0.21}_{-0.21}$  & $0.26^{+0.37}_{-0.16}$  & $0.74^{+0.94}_{-0.40}$ \\ 
  & $<0.14$  & $1.05^{+0.16}_{-0.17}$  & $1.38^{+0.42}_{-0.34}$  & $0.36^{+1.08}_{-0.29}$  & $1.06^{+0.78}_{-0.68}$ \\ 
NMp Sp  & All  & $1.11^{+0.34}_{-0.24}$  & $1.41^{+0.54}_{-0.32}$  & $0.22^{+0.74}_{-0.15}$  & $0.79^{+0.90}_{-0.47}$ \\ 
  & $>0.14$  & $1.11^{+0.28}_{-0.21}$  & $1.34^{+0.43}_{-0.31}$  & $0.19^{+0.52}_{-0.11}$  & $0.66^{+0.99}_{-0.39}$ \\ 
  & $<0.14$  & $1.10^{+0.39}_{-0.25}$  & $1.48^{+0.62}_{-0.32}$  & $0.27^{+0.94}_{-0.21}$  & $0.90^{+0.79}_{-0.50}$ \\   
\hline
\multicolumn{6}{l}{z=3} \\
\hline 
Ms Fs  & All  & $1.01^{+0.12}_{-0.14}$  & $0.99^{+0.39}_{-0.26}$  & $0.97^{+0.69}_{-0.50}$  & $1.09^{+2.02}_{-1.00}$ \\ 
  & $>0.14$  & $0.90^{+0.10}_{-0.10}$  & $0.80^{+0.26}_{-0.08}$  & $0.98^{+0.93}_{-0.41}$  & $1.10^{+1.28}_{-0.84}$ \\ 
  & $<0.14$  & $1.09^{+0.10}_{-0.07}$  & $1.18^{+0.28}_{-0.38}$  & $0.97^{+0.35}_{-0.59}$  & $0.62^{+2.52}_{-0.57}$ \\ 
Ms Fp  & All  & $1.01^{+0.16}_{-0.15}$  & $1.06^{+0.45}_{-0.29}$  & $0.83^{+1.05}_{-0.74}$  & $1.47^{+2.30}_{-1.33}$ \\ 
  & $>0.14$  & $0.97^{+0.13}_{-0.12}$  & $0.99^{+0.19}_{-0.25}$  & $0.65^{+1.59}_{-0.54}$  & $1.32^{+2.28}_{-1.20}$ \\ 
  & $<0.14$  & $1.06^{+0.19}_{-0.14}$  & $1.17^{+0.54}_{-0.33}$  & $0.93^{+0.66}_{-0.87}$  & $1.82^{+1.94}_{-1.68}$ \\ 
Ms Sp  & All  & $0.97^{+0.27}_{-0.21}$  & $1.10^{+0.31}_{-0.35}$  & $0.44^{+0.91}_{-0.41}$  & $1.62^{+2.48}_{-1.30}$ \\ 
  & $>0.14$  & $0.94^{+0.28}_{-0.17}$  & $1.02^{+0.30}_{-0.24}$  & $0.23^{+0.61}_{-0.20}$  & $1.75^{+2.65}_{-0.58}$ \\ 
  & $<0.14$  & $1.00^{+0.26}_{-0.24}$  & $1.12^{+0.45}_{-0.72}$  & $0.78^{+0.93}_{-0.74}$  & $1.60^{+1.65}_{-1.56}$ \\ 
NMp Fs  & All  & $1.02^{+0.16}_{-0.09}$  & $1.10^{+0.35}_{-0.19}$  & $0.71^{+0.76}_{-0.60}$  & $0.60^{+1.27}_{-0.53}$ \\ 
  & $>0.14$  & $0.98^{+0.10}_{-0.10}$  & $1.03^{+0.31}_{-0.12}$  & $0.66^{+0.74}_{-0.28}$  & $0.51^{+0.71}_{-0.45}$ \\ 
  & $<0.14$  & $1.11^{+0.15}_{-0.15}$  & $1.19^{+0.87}_{-0.30}$  & $0.78^{+0.71}_{-0.75}$  & $0.81^{+1.50}_{-0.73}$ \\ 
NMp Fp  & All  & $1.07^{+0.16}_{-0.17}$  & $1.21^{+0.26}_{-0.22}$  & $0.46^{+0.96}_{-0.39}$  & $1.11^{+1.17}_{-0.82}$ \\ 
  & $>0.14$  & $1.05^{+0.13}_{-0.14}$  & $1.19^{+0.21}_{-0.17}$  & $0.22^{+0.54}_{-0.15}$  & $0.95^{+1.23}_{-0.83}$ \\ 
  & $<0.14$  & $1.08^{+0.21}_{-0.21}$  & $1.24^{+0.26}_{-0.33}$  & $0.94^{+0.75}_{-0.81}$  & $1.28^{+1.11}_{-0.89}$ \\ 
NMp Sp  & All  & $1.11^{+0.28}_{-0.24}$  & $1.30^{+0.39}_{-0.26}$  & $0.29^{+1.27}_{-0.24}$  & $1.14^{+1.29}_{-0.93}$ \\ 
  & $>0.14$  & $1.15^{+0.23}_{-0.25}$  & $1.28^{+0.33}_{-0.25}$  & $0.21^{+0.80}_{-0.16}$  & $1.05^{+1.73}_{-0.91}$ \\ 
  & $<0.14$  & $1.08^{+0.34}_{-0.22}$  & $1.32^{+0.49}_{-0.26}$  & $0.63^{+1.47}_{-0.60}$  & $1.23^{+0.99}_{-0.94}$ \\   
\hline
\multicolumn{6}{l}{z=4} \\
\hline 
Ms Fs  & All  & $1.00^{+0.15}_{-0.28}$  & $0.97^{+0.39}_{-0.49}$  & $1.09^{+0.41}_{-0.54}$  & $0.85^{+1.37}_{-0.80}$ \\ 
  & $>0.14$  & $0.91^{+0.17}_{-0.19}$  & $0.85^{+0.35}_{-0.31}$  & $1.13^{+0.75}_{-0.57}$  & $0.88^{+0.92}_{-0.78}$ \\ 
  & $<0.14$  & $1.07^{+0.16}_{-0.15}$  & $1.12^{+0.41}_{-0.64}$  & $1.05^{+0.40}_{-0.77}$  & $0.81^{+2.34}_{-0.77}$ \\ 
Ms Fp  & All  & $1.00^{+0.21}_{-0.14}$  & $1.07^{+0.37}_{-0.40}$  & $0.51^{+0.84}_{-0.48}$  & $2.10^{+2.08}_{-1.98}$ \\ 
  & $>0.14$  & $0.98^{+0.15}_{-0.09}$  & $0.99^{+0.41}_{-0.42}$  & $0.33^{+1.85}_{-0.30}$  & $2.16^{+1.03}_{-2.03}$ \\ 
  & $<0.14$  & $1.01^{+0.25}_{-0.26}$  & $1.12^{+0.36}_{-0.36}$  & $0.73^{+0.55}_{-0.63}$  & $2.07^{+2.80}_{-1.94}$ \\ 
Ms Sp  & All  & $0.98^{+0.37}_{-0.29}$  & $1.11^{+0.37}_{-0.49}$  & $0.29^{+1.82}_{-0.26}$  & $1.36^{+2.12}_{-1.11}$ \\ 
  & $>0.14$  & $0.94^{+0.36}_{-0.24}$  & $1.01^{+0.26}_{-0.46}$  & $0.34^{+2.84}_{-0.32}$  & $1.31^{+1.27}_{-0.91}$ \\ 
  & $<0.14$  & $1.03^{+0.36}_{-0.35}$  & $1.24^{+0.43}_{-0.53}$  & $0.26^{+1.45}_{-0.20}$  & $1.63^{+2.65}_{-1.47}$ \\ 
NMp Fs  & All  & $0.98^{+0.13}_{-0.11}$  & $1.04^{+0.25}_{-0.17}$  & $0.91^{+0.36}_{-0.47}$  & $0.69^{+0.94}_{-0.58}$ \\ 
  & $>0.14$  & $0.92^{+0.16}_{-0.06}$  & $0.95^{+0.22}_{-0.11}$  & $0.92^{+0.35}_{-0.40}$  & $0.48^{+1.36}_{-0.38}$ \\ 
  & $<0.14$  & $1.04^{+0.11}_{-0.10}$  & $1.12^{+0.31}_{-0.21}$  & $0.90^{+0.32}_{-0.54}$  & $0.89^{+0.57}_{-0.77}$ \\ 
NMp Fp  & All  & $1.05^{+0.17}_{-0.16}$  & $1.19^{+0.45}_{-0.24}$  & $0.55^{+0.83}_{-0.44}$  & $1.08^{+2.05}_{-0.73}$ \\ 
  & $>0.14$  & $1.00^{+0.19}_{-0.12}$  & $1.11^{+0.42}_{-0.18}$  & $0.46^{+0.68}_{-0.40}$  & $0.76^{+1.28}_{-0.41}$ \\ 
  & $<0.14$  & $1.09^{+0.17}_{-0.17}$  & $1.25^{+0.48}_{-0.27}$  & $0.79^{+0.68}_{-0.62}$  & $1.59^{+2.05}_{-1.25}$ \\ 
NMp Sp  & All  & $1.12^{+0.64}_{-0.25}$  & $1.24^{+0.33}_{-0.38}$  & $0.47^{+4.68}_{-0.40}$  & $1.45^{+1.98}_{-1.29}$ \\ 
  & $>0.14$  & $1.13^{+0.48}_{-0.18}$  & $1.21^{+0.25}_{-0.27}$  & $0.43^{+5.74}_{-0.39}$  & $1.19^{+2.02}_{-1.04}$ \\ 
  & $<0.14$  & $1.11^{+0.87}_{-0.32}$  & $1.29^{+0.35}_{-0.48}$  & $0.55^{+4.19}_{-0.46}$  & $1.64^{+1.94}_{-0.95}$ \\ 
\hline
\multicolumn{6}{l}{z=5} \\
\hline 
Ms Fs  & All  & $0.97^{+0.19}_{-0.25}$  & $0.98^{+0.31}_{-0.55}$  & $1.01^{+0.42}_{-0.91}$  & $1.28^{+2.73}_{-0.62}$ \\ 
  & $>0.14$  & $0.87^{+0.10}_{-0.34}$  & $0.80^{+0.24}_{-0.43}$  & $0.86^{+0.41}_{-0.71}$  & $1.27^{+1.73}_{-0.63}$ \\ 
  & $<0.14$  & $1.07^{+0.16}_{-0.12}$  & $1.12^{+0.31}_{-0.41}$  & $1.09^{+0.35}_{-1.04}$  & $1.28^{+3.65}_{-0.33}$ \\ 
Ms Fp  & All  & $1.05^{+0.23}_{-0.34}$  & $1.04^{+0.47}_{-0.56}$  & $0.73^{+1.21}_{-0.67}$  & $2.39^{+2.56}_{-1.62}$ \\ 
  & $>0.14$  & $1.04^{+0.31}_{-0.40}$  & $1.04^{+0.39}_{-0.49}$  & $0.48^{+1.46}_{-0.43}$  & $2.08^{+2.36}_{-1.21}$ \\ 
  & $<0.14$  & $1.06^{+0.19}_{-0.22}$  & $1.06^{+0.51}_{-0.58}$  & $0.91^{+1.03}_{-0.72}$  & $2.73^{+2.96}_{-2.54}$ \\ 
Ms Sp  & All  & $0.96^{+0.33}_{-0.35}$  & $1.07^{+0.40}_{-0.78}$  & $0.14^{+1.51}_{-0.12}$  & $2.02^{+4.61}_{-1.51}$ \\ 
  & $>0.14$  & $0.97^{+0.19}_{-0.33}$  & $1.06^{+0.31}_{-0.51}$  & $0.07^{+0.98}_{-0.05}$  & $1.66^{+5.87}_{-1.13}$ \\ 
  & $<0.14$  & $0.93^{+0.51}_{-0.34}$  & $1.07^{+0.52}_{-0.78}$  & $0.34^{+2.39}_{-0.33}$  & $2.17^{+3.76}_{-1.67}$ \\ 
NMp Fs  & All  & $0.96^{+0.19}_{-0.18}$  & $1.00^{+0.32}_{-0.47}$  & $1.01^{+0.47}_{-0.40}$  & $0.54^{+1.81}_{-0.37}$ \\ 
  & $>0.14$  & $0.91^{+0.11}_{-0.16}$  & $0.95^{+0.13}_{-0.28}$  & $0.93^{+0.50}_{-0.25}$  & $0.32^{+0.96}_{-0.17}$ \\ 
  & $<0.14$  & $1.02^{+0.21}_{-0.18}$  & $1.04^{+0.44}_{-0.60}$  & $1.02^{+0.55}_{-0.97}$  & $0.87^{+2.66}_{-0.64}$ \\ 
NMp Fp  & All  & $1.08^{+0.20}_{-0.14}$  & $1.28^{+0.50}_{-0.36}$  & $0.60^{+0.84}_{-0.56}$  & $0.73^{+2.12}_{-0.63}$ \\ 
  & $>0.14$  & $0.99^{+0.24}_{-0.16}$  & $1.10^{+0.37}_{-0.22}$  & $0.43^{+0.87}_{-0.33}$  & $0.49^{+1.74}_{-0.39}$ \\ 
  & $<0.14$  & $1.14^{+0.16}_{-0.12}$  & $1.47^{+0.45}_{-0.44}$  & $0.63^{+0.83}_{-0.62}$  & $0.94^{+2.94}_{-0.80}$ \\ 
NMp Sp  & All  & $1.11^{+0.65}_{-0.23}$  & $1.30^{+0.49}_{-0.35}$  & $0.36^{+1.62}_{-0.31}$  & $0.92^{+1.94}_{-0.73}$ \\ 
  & $>0.14$  & $1.21^{+1.16}_{-0.19}$  & $1.27^{+0.40}_{-0.41}$  & $0.54^{+10.26}_{-0.46}$  & $0.52^{+1.98}_{-0.37}$ \\ 
  & $<0.14$  & $1.00^{+0.38}_{-0.23}$  & $1.30^{+0.80}_{-0.32}$  & $0.21^{+1.06}_{-0.18}$  & $1.34^{+2.12}_{-1.05}$ \\ 
\enddata
\end{deluxetable}

\end{document}